\documentclass[%
 aip,
 amsmath,amssymb,
preprint,%
author-year,%
]{revtex4-1}
\usepackage{lineno}
\usepackage{graphicx}
\usepackage{color,soul}
\usepackage{multirow}
\usepackage{placeins}
\usepackage{soul}
\usepackage{subcaption}
\usepackage{hyperref}

\draft 

\begin{document}

\fbox{ 
\parbox{0.95\textwidth}{
This article may be downloaded for personal use only. Any other use requires prior permission of the author and AIP Publishing. This article appeared in \emph{Physics of Fluids} and may be found at \href{ https://doi.org/10.1063/5.0088000}{https://doi.org/10.1063/5.0088000}.
Please cite as: \emph{Bultreys, T., et al (2022). X-ray tomographic micro-particle velocimetry in porous media. Physics of Fluids, 34(4), 042008.} }
}

\title[]{X-ray Tomographic Micro-Particle Velocimetry in Porous Media}
\author{T.Bultreys }
\email[Corresponding author, ]{Tom.Bultreys@UGent.be}

\author{S. Van Offenwert}%
\affiliation{ 
Centre for X-ray Tomography, Ghent University, Proeftuinstraat 86, 9000 Ghent, Belgium
}%
\affiliation{ 
Department of Geology, Ghent University, Krijgslaan 281, 9000 Ghent, Belgium 
}%

\author{W. Goethals}
\author{M.N. Boone}
\affiliation{ 
Centre for X-ray Tomography, Ghent University, Proeftuinstraat 86, 9000 Ghent, Belgium
}%
\affiliation{%
Department of Physics and Astronomy, Ghent University, Proeftuinstraat 86, 9000 Ghent, Belgium
}%
\author{J. Aelterman}
\affiliation{ 
Centre for X-ray Tomography, Ghent University, Proeftuinstraat 86, 9000 Ghent, Belgium
}%
\affiliation{%
Department of Physics and Astronomy, Ghent University, Proeftuinstraat 86, 9000 Ghent, Belgium
}
\affiliation{%
Department of Telecommunications and information processing – imec, Ghent University, Sint-Pietersnieuwstraat 41, 9000 Ghent, Belgium
}%

\author{V. Cnudde}
\affiliation{ 
Centre for X-ray Tomography, Ghent University, Proeftuinstraat 86, 9000 Ghent, Belgium
}%
\affiliation{ 
Department of Geology, Ghent University, Krijgslaan 281, 9000 Ghent, Belgium 
}%
\affiliation{ 
Department of Earth Sciences, Environmental Hydrogeology, Utrecht University, Princetonlaan 8A, 3584 CS Utrecht, The Netherlands
}%
\date{\today}

\begin{abstract}
Fluid flow through intricate confining geometries often exhibits complex behaviors, certainly in porous materials, e.g. in groundwater flows or the operation of filtration devices and porous catalysts. However, it has remained extremely challenging to measure 3D flow fields in such micrometer-scale geometries. Here, we introduce a new 3D velocimetry approach for optically opaque porous materials, based on time-resolved X-ray micro-computed tomography (µCT). We imaged the movement of X-ray tracing micro-particles in creeping flows through the pores of a sandpack and a porous filter, using laboratory-based µCT at frame rates of tens of seconds and voxel sizes of 12 µm. For both experiments, fully three-dimensional velocity fields were determined based on thousands of individual particle trajectories, showing a good match to computational fluid dynamics simulations. Error analysis was performed by investigating a realistic simulation of the experiments. The method has the potential to measure complex, unsteady 3D flows in porous media and other intricate microscopic geometries. This could cause a breakthrough in the study of fluid dynamics in a range of scientific and industrial application fields.  

\end{abstract}


\maketitle 

\section{Introduction}\label{Introduction}
Fluid dynamics in porous materials play an important role in nature and in industry, e.g. groundwater flow in aquifers \citep{Mercer1990}, gas-brine flow in geological energy or carbon storage reservoirs \citep{Mouli-Castillo2019, Bui2018}, and the performance of filtration devices, fuel cells and catalysts \citep{Miele2019, Mularczyk2020}. The intricate pore geometries in such materials can lead to complex phenomena, particularly during  {solute and colloid transport \citep{Zhang2021, Haffner2020, Russell2021}}, multiphase flows \citep{Blunt2017, Singh2019b} and non-Newtonian flows  {\citep{An2022}}. While experiments on simplified (often 2D) model geometries give valuable insights on flow behavior in the confinement of generic pore walls \citep{Primkulov2019, Lenormand1983, Holtzman2016, Datta2014}, the physical interactions in the complex 3D pore networks encountered in many applications remain difficult to probe. This is important as highly irregular pore geometry and connectivity are known to influence the emerging behavior at the macro-scale in a non-trivial way, due to the non-linearity of many flow processes in porous media \citep{Mascini2021a, Ling2017, McClure2021}. Recent pore-scale numerical simulation methods can - to a certain extent - be applied to study these porous media, but often still come with significant uncertainties on the incorporated physical assumptions and materials properties \citep{Zhao2019, Ye2019}. Furthermore, such methods are in many cases severely restricted by either the computational time, domain size or accuracy. There is thus an important need for \emph{in-situ} experimental measurements to study 3D porous media flows at the scale of the flow-confining pore geometries (nm to mm). 

For the wider field of experimental fluid mechanics, the introduction of methods to measure 3D flow and pressure fields has been a turning point, as reviewed by \citet{Discetti2018}. However, this has not been applicable to a majority of porous materials of interest to the research community, due to the optical opacity of these materials. Most flow field measurements are based on optical particle velocimetry, using visible light to image the movement of flow-tracing particles in (index-matched) fluids over time. With micro-particles and microscopes, this principle can be used to measure micron-scale flow fields in transparent 2D micromodels \citep{Roman2015, Zarikos2018a} and even in optically transparent 3D porous media, using multi-camera set-ups \citep{Schanz2016}, astigmatic optics \citep{Franchini2019} or confocal microscopy \citep{Datta2013, Datta2014a}. However, these techniques are inherently unsuited for optically opaque - and thus most - porous materials. An alternative method is to measure fluid propagators using (pulsed-field gradient) magnetic resonance imaging \citep{Gladden2013}. While having several advantages, including not requiring tracers, this method has only recently started to reach the required micron-scale spatial resolutions \citep{DeKort2019}. Several hours are required to measure a single flow field at this resolution, restricting its applicability to static flow fields.

In this paper, we introduce a 3D micro-particle velocimetry method for porous media by leveraging the penetrating power of X-rays. Contrary to previous methods, the approach is applicable to tortuous, spatially varying 3D flow fields common in porous media, and can be extended to unsteady flows. Prior approaches to X-ray based particle velocimetry started with 2D, radiography-based measurements, which did not yield 3D information \citep{Lee2003}. This was followed by methods that reconstructed 3D flow fields from correlations within radiography sequences, taken from different viewing angles \citep{Fouras2007, Dubsky2012, Baker2018}. While high particle velocities could be measured because radiographs can be acquired mere milliseconds apart, these methods have only been applied to fairly homogeneous flows in e.g. a blood vessel, and it is unclear how well suited their reconstruction algorithms are to complex flow fields in porous media. The alternative method we adopt here is X-ray micro-computed tomography (µCT), an inherently 3D, non-destructive and micrometer-scale technique \citep{Cnudde2013, Wildenschild2013}, to reconstruct a time series of 3D images of flow-tracing particles. The challenge is to precisely resolve the locations of the tracer particles at a sufficiently high frame rate. For their motion to be representative of the flow, these particles should be small and close in mass density to the liquid to negate inertial and gravitational effects. However, this tends to negatively affect the particles' visibility in X-ray imaging. Furthermore, µCT imaging typically takes tens of minutes to acquire a 3D image, which is too slow to track the particle movement. Time resolutions on the scale of (tens of) seconds have only become possible at synchrotrons a few years ago \citep{Berg2013}, and even more recently in laboratory-based µCT scanners \citep{Bultreys2015f}.

Very recently, \citet{Makiharju2022} provided a proof-of-concept that flow tracer particles (60 µm large silver-coated hollow glass spheres) in a cylindrical tube could be visualized with laboratory-based µCT at frame rates on the order of seconds. Here, we present the first successful µCT-based particle velocimetry measurements of creeping single-phase flow in porous media, namely a sandpack and a sintered glass filter. Our method yields fully 3D, 3-component velocity fields, by tracing the movement of thousands of individual silver-coated micro-spheres with a mean diameter of approximately 20 µm. The measurements were performed using a laboratory-based µCT scanner at a voxel size of 12 µm and an acquisition time of 70s per scan, with a total measurement time of 30 to 45 minutes. 

In the following, we first introduce the basic concepts of particle tracking velocimetry in Section \ref{Intro XPTV}. The experimental workflow is described in Section \ref{Experiments}. We used a Lagrangian Particle Tracking approach to identify individual particle trajectories in the image, and interpolated the resulting velocity data points to find velocity fields, as explained in Section \ref{Data analysis}. The method was validated on a realistic numerical simulation of an imaging experiment, which provided ground-truth data to validate particle locations and velocities. The generation of this dataset is treated in Section \ref{Validation}. The results of the experiments and the validation are discussed in Section \ref{Section Results}.


\section{Materials and methods}\label{Materials and methods}
\subsection{Introduction to particle tracking velocimetry}\label{Intro XPTV}
Particle velocimetry (PV) methods work by computing the displacement of flow tracer particles in a time series of images. Before introducing specific approaches, it is useful to discuss following general considerations when selecting or applying these methods:
\begin{itemize}
    \item The \emph{sampling density} of the resulting velocity field is the density of the cloud of points in which particles were detected and velocities could thus be measured. This has the potential to improve with longer measurement time or denser particle seeding, as well as with the resolution of the particle images.
    \item The \emph{measurement time} refers to the time needed to acquire all the data to reconstruct a velocity field. This determines whether changes in (unsteady) flows over time can be measured.
    \item The \emph{frame interval} is the time interval between consecutive particle images (frames). To track the paths of fast-moving particles, this time interval needs to be small enough. 
    \item The \emph{acquisition time} is the time to acquire a single frame. This needs to be small enough to accurately measure the particle positions, as their motion would otherwise cause blurring and other image artifacts. In optical imaging, the frame interval is larger than or equal to the acquisition time, but in our method this is not necessarily the case, as discussed in Section \ref{Experiments}.
    \item The \emph{particle seeding concentration} is the amount of particles in a unit volume of liquid. Higher concentrations can improve the spatial or temporal resolution, but come at the cost of higher computational complexity. In porous media, high seeding concentrations may also induce pore clogging.
    \item The \emph{tracer fidelity} refers to the need for good flow tracers to follow the flow lines of the liquid, rather than be significantly influenced by inertia or gravity. Particles should thus have a small Stokes number and a small ratio of gravitational settling velocity to flow velocity \citep{Melling1997}. This depends on the liquid’s viscosity and on the size and material density of the particles.
\end{itemize}

There are two main classes of approaches to PV. The first and most well-known, particle image velocimetry (PIV), yields a flow field from as little as two snapshots of the particles, by dividing the images into small windows that typically contain multiple particles, and investigating correlations between these windows in consecutive time steps \citep{Raffael2018}. Particle tracking velocimetry (PTV, also called Lagrangian Particle Tracking), on the other hand, identifies the locations of individual particles as they travel throughout many images. PIV tends to have a better temporal resolution because it requires fewer particle images and deals better with high seeding concentrations, while PTV yields more precisely localized velocity information, as well as Lagrangian properties of the flow \citep{Ouellette2006}. 

In this work, we employ PTV, for two main reasons. First, we aim to measure flows in geometries bound by irregular pore walls, which means that the spatial discretization of PIV into interrogation windows may cause issues. Second, out of concern to avoid significant pore clogging by particle straining \citep{Molnar2015}, we have kept the seeding concentrations relatively low - making PTV the more favorable option. 

\subsection{Experiments} \label{Experiments}
In the following, we first describe how the flow experiments were performed and then provide a detailed description of the tracer particle suspension used in these experiments, which consisted of silver-coated hollow glass spheres in a highly viscous glycerol-water mixture.

\subsubsection{Flow experiments} \label{Flow experiments}
\begin{figure}[htb]
\centering
\includegraphics[width=0.6\linewidth]{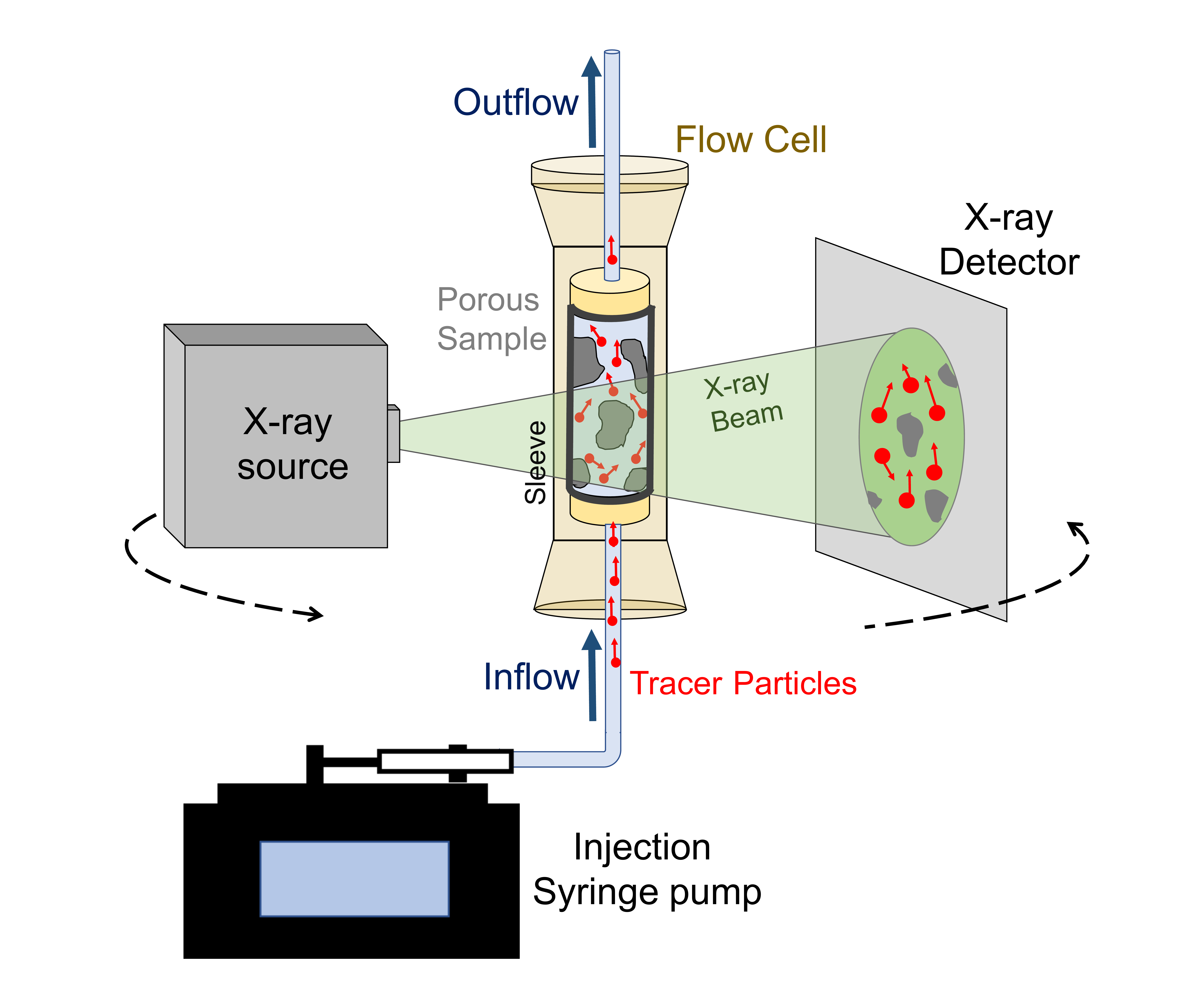}
\caption{A schematic of the experimental setup used in this work (not drawn to scale). The samples were 4 mm in diameter and 10 to 20 mm high.}
\label{Figure: setup}
\end{figure}

We present velocimetry experiments on two porous samples: a sand pack and a porous glass filter. The first sample was a construction-grade sand used in mortars (HUBO, Belgium), sieved to retain the fraction of grains between 500 and 710 µm. The grains were poured into a viton sleeve of 4 mm diameter mounted in a flow cell, to a sample height of approximately 20 mm. The second was a cylindrical sintered glass filter of 4 mm diameter and 10 mm height, with nominal pore sizes between 160 and 250 µm (ROBU P0, Germany), in a viton sleeve. Image-based estimates of the porosity and mean pore (throat) sizes of the samples are listed in Table \ref{Table fluid particle properties}. Both samples were mounted into a vertically-oriented, X-ray transparent Hassler-type flow cell (RS Systems, Norway). We avoided flow from bypassing the sample by pressurizing the sleeve around the samples with a confining pressure of 2 MPa. The liquids were injected from the bottom to the top of the samples. The flow cell was mounted on the Environmental µCT scanner (EMCT) at Ghent University's Centre for X-ray Tomography: a fast-scanning system which does not rotate the sample like most µCT scanners, but instead rotates a source-detector system on a gantry around it. A detailed description of the scanner and its application to fast imaging can be found in \citet{Dierick2014} and \citet{Bultreys2015f}. A schematic of the setup is shown in Figure \ref{Figure: setup}.

Before the velocimetry experiment, the samples were saturated with the unseeded glycerol-water liquid by flooding more than 50 pore volumes of liquid through the sample at a high flow rate, namely a Darcy velocity (fluid flux) of 2 mm/s, to mobilize trapped air. Then, a high-quality µCT scan at 6 µm voxel size was made of the field-of-view of the experiment: a section of the sample near the inlet, approximately 6.3 mm high and containing its full diameter (2200 projections, 110 kV accelerating voltage, 8W X-ray power, 550 ms integrated exposure time per projection). 

To start the experiment, the tracer particle suspension was drawn up in high-precision glass syringes (Hamilton GasTight Syringe models 1001 and 1002, USA) and injected into the sample at the flow rates listed in Table \ref{Table fluid particle properties}, using a Harvard PHD Ultra syringe pump (Harvard, USA). The tracers were injected from the bottom-side of the sample, via PTFE tubing. The imposed  {constant} volumetric flow rates in the experiments were set to arrive at an estimated average interstitial velocity around 1 voxel per scanner rotation (0.5 voxels/frame  {due to the interleaved reconstruction procedure explained below}). These rates were at least 10 times larger than the minimum setting of the syringe pump, ensuring smooth flow. The pump accuracy and reproducibility were respectively 0.25\% and 0.05\%. After assessing that the tracer particles were present in the sample by radiography, continuous µCT acquisitions of either 30 or 40 back-to-back rotations were started, at 70s per full rotation and 11.8 µm voxel size (700 projections/rotation, 100 ms exposure time, 60 kV accelerating voltage, 8W X-ray power). 

After the experiment, the data were reconstructed into time series of 3D frames using a filtered back projection algorithm (Tescan XRE, Belgium), taking 700 projections per frame and 350 projections in between each two frames. This resulted in an \lq\lq{}interleaved\rq\rq{} time series with a frame interval of 35 s and an acquisition time of 70 s. An intuitive way to understand the resulting data is to compare it to an interleaved stream of images from two cameras with staggered trigger time, so that the exposures of each two consecutive frames overlap by 50\%. This reduced the maximum distance traveled by a particle between two consecutive frames, which was beneficial for the particle tracking algorithm.  

Minor amounts of particle retention  {were found during visual inspection of cross-sectional slices} of the reconstructed images  {(e.g. Figure \ref{cross-section grey value video} in the Appendix)}, and were deemed to have a limited effect on the velocimetry results as pore clogging was negligible. However, this issue did cause several failed experimental trials during the method's development. Its effects typically became severe after pumping a few pore volumes (i.e. tens of µL) of the tracer-seeded liquid through the imaged part of the sample. During both experiments, which took respectively 46 and 35 minutes for the sandpack and the porous glass filter, only 1.5\,µL of liquid (approximately 5\% of the imaged pore volume) was pumped. Carefully timing the arrival of the tracers with the start of the acquisition was thus key. This was achieved by inspecting the sample with radiography during particle delivery while remotely controlling the pump. Note that the risk of tracer retention can be significantly reduced by decreasing their particle size compared to the pore {-throat} size.  {While a systematic study of the minimum pore throat-tracer size ratio needed to perform velocimetry experiments is out of the scope of this study, preliminary tests did {show} significant clogging in samples where the mean pore-throat size was close to the mean tracer particle diameter. As a reasonable working hypothesis, we assumed that the maximum tracer size (here: 60 µm, see Section \ref{Particle-liquid system}) should be smaller than typical pore-throat size of the main flow paths, which we estimated by the mean pore-throat size in Table \ref{Table fluid particle properties} }. Successfully resolving smaller particles could be achieved by increasing the spatial resolution of the images  {without impacting the image quality or the temporal resolution} (e.g. using synchrotron µCT).

\begin{table}[htb]
\caption{Key experimental properties for the velocimetry experiments. Mean pore and throat sizes were based on the open-source pore network extraction algorithm PNExtract \citep{Raeini2017}.  Glycerol-water properties were based on tabulated data \citep{Segur1951, Takamura2012}}.
\label{Table fluid particle properties}
\begin{tabular}{|l|cc|}
\hline
\textbf{Experiment}                              & \, \textbf{Sand pack}\, & \, \textbf{Porous glass}\, \\ \hline
\textbf{Sample} & & \\
Mean pore size (µm)                     & 172       & 163           \\
Mean throat size (µm)                   & 95        & 78           \\
Image-based porosity (\%)               & 34.1      & 27.5           \\\cline{1-3}
\textbf{Tracer suspension} & & \\
Glycerol concentration (wt\%) in water\,           & 95        & 93           \\
Viscosity (cP)                          & 523       & 367            \\
Liquid mass density (g/ml)              & 1.247     & 1.242           \\
Seeding concentration (mg/g)            & 8         & 12           \\\cline{1-3}
\textbf{Flow properties} & & \\
Flow rate (nl/min)                      & 33        & 44           \\
Interstitial velocity (nm/s)            & 128        & 212            \\
Interstitial velocity (voxel/ {time frame})      & 0.38        & 0.62            \\
Gravitational settling for mean tracer size (nm/s) \, &  {59}        &   {85}            \\\cline{1-3}
\textbf{Imaging settings} & & \\
Number of (interleaved) time frames                    & 80        & 60            \\
µCT voxel size (µm)                & \multicolumn{2}{c|}{11.8}  \\
µCT image size (voxels)            & \multicolumn{2}{c|}{658 x 658 x 539}            \\
Acquisition time 3D images (s)           & \multicolumn{2}{c|}{70}    \\
Frame interval 3D images (s)             & \multicolumn{2}{c|}{35}            \\
\hline
\end{tabular}
\end{table}


\subsubsection{Particle-liquid system} \label{Particle-liquid system}
\begin{figure}[htb]
\centering
\includegraphics[width=0.5\linewidth]{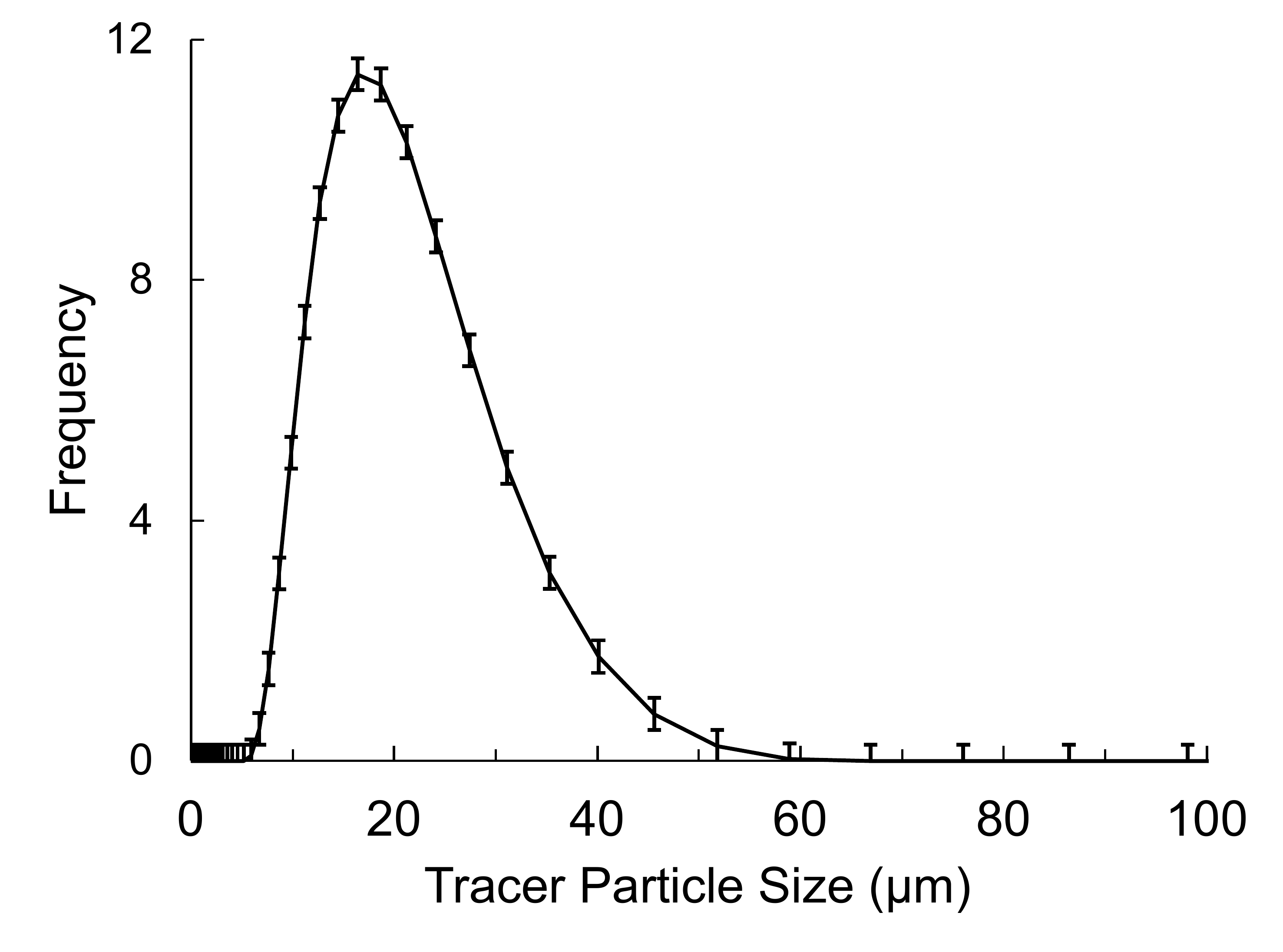}
\caption{We measured the size distribution of the silver coated hollow glass spheres that were used as tracer particles in the experiments with laser diffraction. The error bars reflect the standard deviation on 5 measurements.}
\label{Particle size distribution}
\end{figure}

The flow tracing particles were hollow glass microspheres with a nominal particle size range between 5 and 22 µm and a 250 nm thick silver coating, resulting in a particle mass density of 1.4 g/ml (Cospheric, USA). We measured the particle size distribution of the tracer with a laser diffraction particle sizer (Malvern MasterSizer 3000, UK), indicating  {a mean size of 19.3 µm} and the occurrence of larger sizes than the nominal range, up to 60 µm (Figure \ref{Particle size distribution}). For the velocimetry experiments, the tracer particles were suspended in high-viscosity mixtures of glycerol and water, with 93 - 95 weight percent glycerol. The silver coating had a high X-ray attenuation coefficient due to its high atomic number, providing a beneficial contrast with the liquid in the images, contrary to what can be expected from traditional micro-velocimetry particles such as polyethylene microspheres. 

The high-viscosity liquids caused a strong drag force on the particles, preventing that their inertia would cause deviations from the liquid's flow lines (Stokes numbers were of the order of 10$^{-10}$ or smaller). Furthermore, the viscosity reduced the speed of gravitational settling. The terminal sinking velocity for a sphere with radius R can be calculated using Stokes' law:
\begin{equation}
    v = \frac{2}{9} \frac{\rho_{s} - \rho_{f}}{ \mu } g R^2
\end{equation}
with $\rho_{s}$ and $\rho_{f}$ the mass densities of respectively the particle and the fluid, $\mu$ the viscosity, and $g$ the gravitational acceleration (the values of these material properties are listed in Table \ref{Table fluid particle properties}). In the experiments, the estimated interstitial velocity (based on the imposed volumetric flow rate) was two to three times higher than the settling velocity  {for the mean particle size} (Table \ref{Table fluid particle properties}).  {In the respective experiments, particles larger than 28.5 µm and 27.5 µm were expected to have gravitational velocities equal to or larger than the flow velocity. Note that the largest of these particles may never reach the sample through the vertical tubing below the flow cell}. Nevertheless, gravitational settling may thus still lead to an underestimation in the vertical component (along the Z-axis) of the velocity field.  {However, this} issue can be reduced by using smaller or lighter particles, or faster interstitial velocities, as the imaging methods become more powerful. 

The particles were added to the liquid with seeding concentration of 8 or 12 mg/g (estimated 5.5 or 8.3 million particles per ml of liquid,  {which translates to 0.009 or 0.014 particles per voxel}).  {This was based on trial-and-error, and may be further adapted: increasing the seeding concentration may lead to lower measurement times, while decreasing it may reduce clogging due to jamming effects in cases where this causes issues}. The suspension was first vigorously stirred, then treated with an ultrasonic homogenizer (Hielscher UP50H, Germany) for 5 minutes and  placed in an ultrasonic bath (Bandelin Sonorex TK52, Germany) for 10 minutes to disperse the particles and remove air bubbles, respectively. The particle dispersion was slightly less succesful in the  {sand pack experiment, likely due to technical issues with the ultrasonic equipment}.

\subsection{Image processing and velocimetry analysis}\label{Data analysis}
\subsubsection{Particle tracking algorithm} 
To track trajectories of individual particles in the time series of 3D µCT frames acquired during the experiment, particles were first detected in each time step image, and these detections were then associated between time steps to result in particle tracks. A multitude of methods to do this are compared in \citet{Chenouard2014}. In this work, we used the Crocker and Grier method implemented in the open-source Python package TrackPy \citep{Crocker1996, VanDerWel2022}. First, the background was subtracted from the time series images by registering and  {downsampling} the high-quality pre-scans to the time series in Avizo (Thermo Fisher, France). Then, potential particle locations were identified as local grey value maxima in the background-subtracted images using TrackPy. Any particle detection outside of the pore space was removed by masking the experimental images with a segmentation of the pore space from the registered pre-scan (made using simple grey value thresholding in Avizo). The locations of local grey-value maxima were adopted as particle locations if no voxels within a \emph{minimum particle separation distance} of 4 voxels had a higher grey value than that maximum, and if they lied within the brightest 2 percent of grey values in the pore space. These values were set by visual inspection of the particle identifications in the images. The particle locations were then refined to a sub-voxel accuracy by calculating the brightness-weighted centroid of the voxels in a neighbourhood around the peak value. Finally, noisy particle detections with very low brightness-weighted mass were removed.  

A location-prediction based nearest-neighbour algorithm was used to identify which particle observation in a certain frame most likely corresponded to a certain observation in the previous frame \citep{Crocker1996}. For each particle, the local velocity was estimated from 3 prior time steps to predict its new location in each time step (the displacement was initialized to a user-defined value in the first time step). The observation nearest to this predicted location was then taken as the particle's new location. To keep the calculations tractable, a maximum search range of 6 voxels around the predicted location was set to limit the amount of potential matches. The association step would therefore become more challenging if the average displacement between time frames were to be larger or if the particle seeding density were to be increased. It should be noted that more advanced methods than the nearest-neighbour approach have been developed, relying on e.g. multiple hypothesis tracking or Kalman filtering \citep{Jaqaman2008, Chenouard2014, Godinez2015}. These methods are computationally more demanding, but may be of benefit in experiments with larger particle displacements or seeding densities than the ones presented here. The current analysis took less than 1 hour to treat a full experimental data set, running on the CPU of a moderately-sized work station (Intel Core i7-8700 with 64 GB RAM).

\subsubsection{Velocity field interpolation  {and comparison to computational fluid dynamics}} \label{velocity field interpolation and CFD}
After identifying particle trajectories, those that were only a few time steps long were removed, as these typically contained noisy detections. In both experiments, a minimum trajectory span of 20 frames was set. Particle velocity vectors were calculated for each remaining particle track using a centered finite difference approach. The resulting cloud of velocity vectors was then linearly interpolated on a grid  {with the same} voxel size as the experimental  {time step images} (using SciPy), to find the 3D field of all three velocity components. To take into account that velocities should be zero in the solid material during the interpolation, zero-velocity points were added in a randomly selected fraction of the pore wall voxels (2.5\%). Higher-order interpolation and adding zero-points in all boundary voxels was computationally prohibitive as the interpolation code remained to be optimized for computational efficiency. 

 {To evaluate the measured velocity fields, we performed a cross-validation to a computational fluid dynamics (CFD) approach to calculate the velocity fields in the pore space geometry. We used an open-source finite volume solver based on OpenFOAM, from \citet{Raeini2022}. The solver performed finite-volume calculations on a hexahedral mesh extracted from the segmented pre-scan of the pore space, with constant-pressure and zero-velocity-gradient boundary conditions at the in- and outlet. We used the standard  {code} provided by \citet{Raeini2022}. Note that the CFD result should not be considered as ground truth in this comparison, and differences compared to the experiments may result from both measurement errors and numerical errors \citep{Saxena2017}.}



\subsection{Simulated  {µCT} data sets for method validation} \label{Validation}
To validate the imaging and particle tracking workflow, we generated simulated µCT datasets based on ground-truth particle locations. This was done by computer-generating spatial distributions of analytical spheres with specified diameters and velocities, and then simulating radiographs by tracing rays from a point source to a detector array through these digital samples, with the tracer particle locations being updated in each radiograph. This way, simulated scans contained realistic geometrical deformation and motion artifacts, as particle locations changed in each radiograph (note that particle motion within individual radiographs was negligible as these are typically acquired on the ms time scale). 

The validation data was meant to mimic a velocimetry experiment in a porous medium as closely as possible. To this end, we took the segmented pore geometry of the Porous Glass experiment (Section \ref{Flow experiments}) as input, and determined its CFD-based velocity field (see Section \ref{velocity field interpolation and CFD}). The velocity field was scaled to an average velocity magnitude of 1 voxel per 360\textdegree\ scan. Then, to reflect the particle seeding of an incompressible flow, the initial positions of the simulated \lq\lq{}tracer\rq\rq{} spheres were chosen randomly in the the pore space (staying clear of the in- and outlet boundaries), with sphere radii drawn from the experimentally measured tracer particle size distribution (Figure \ref{Particle size distribution}). The tracer seeding density was tuned to approximate the Gaussian-like distribution of inter-particle distances from the experiment, and was therefore also set to zero in regions with very low velocities (lower than 5\% of the maximum). Next, the locations of these particles were calculated as they moved through the CFD-based velocity field for 4900 time steps of 100 ms, using a 4th order Runge-Kutta integration. These were the ground-truth locations for the simulated µCT data set, regardless of potential numerical errors in their determination (the purpose of the calculation being only to create ground-truth trajectories with a realistic complexity).  {This way, the particle positions were calculated for each radiograph time step in 7 consecutive µCT scans of 360\textdegree\ and 700 radiographs each, matching the experimental acquisition. Each of these radiographs was then calculated by raytracing using the in-house developed CTRex code \citep{Heyndrickx2020, DeSchryver2018}.} Poisson noise was added on the radiographs to match the noise level in the experiment. Finally, the simulated dataset was reconstructed with filtered back-projection. Contrary to the experiments, there was no frame interleaving in the reconstruction of the simulated data sets, to aid the interpretation and maximize the generality of the results. The simulated acquisition time and frame interval were both 70\,s. The result is a time series of simulated µCT images of tracer particles moving through a porous medium with exactly known trajectories, in order to investigate the errors expected in the experimental data.

\section{Results} \label{Section Results}
\subsection{Experimental results} \label{Experimental results}

Visual inspection of cross-sections through the imaged porous media confirmed that tracer particles were visible as bright spots of a few voxels in diameter, which moved slowly and smoothly through the pore space (videos in Appendix). In the sand pack, tracer particles appeared slightly larger, which may be due to particle aggregates being more difficult to disperse in the higher viscosity liquid used for this experiment. In regions with high flow rates, motion artifacts appeared to be present in the form of slightly blurred particle shapes elongated in the direction of motion, rather than as severe corkscrew-shaped artifacts and streaks which would occur in the case of large movements. The observed deformation did not necessarily cause issues in the particle localization, as this was based on the particles' grey value centroids. This will be investigated in more detail in Section \ref{Validation results}. 

\begin{figure}[htb]
\centering
\includegraphics[width=0.8\linewidth]{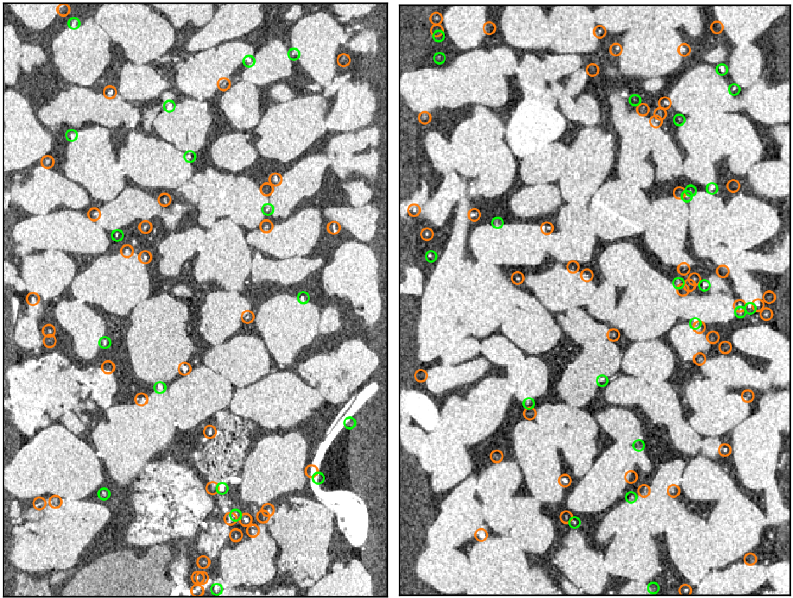}
\caption{This image illustrates the frame-by-frame particle detection algorithm, applied to one 3D frame of the sandpack (left) and the glass filter (right) experiment. Detections that fall in the pictured cross-sectional slice are indicated in green, detections in a directly neighbouring slice are indicated in orange.}
\label{Grey value images}
\end{figure}

Frame-by-frame particle detection yielded $2393 \pm 86 $ particles per frame in the sand pack experiment, and  $5581 \pm 136 $ in the porous glass. Example slices through the 3D data with annoted particle detections are shown in Figure \ref{Grey value images}. In the ideal scenario that there is no particle agglomeration, the selected seeding concentrations would yield on average 1 particle in a cube with a side length of 4.5 voxels (sand pack) or 4.1 voxels (porous glass). The measured average distance between neighboring particles was larger than expected due to particle agglomeration and (mainly small) particles going undetected: 9.7 voxels and 8.0 voxels in the respective experiments. 

\begin{figure}[htb]
\centering
\includegraphics[width=0.9\linewidth]{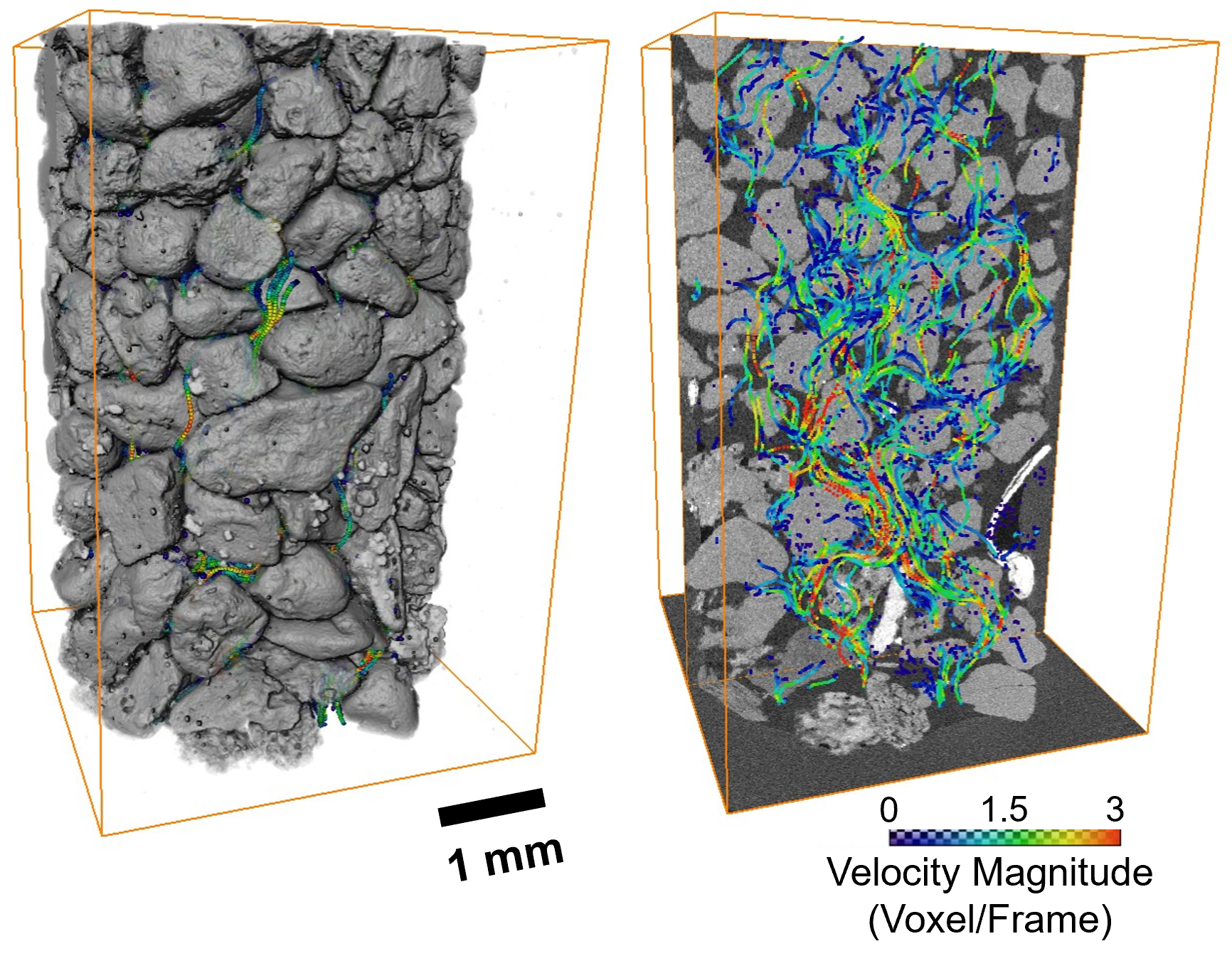}
\caption{Experimentally measured trajectories of tracer particles in the sand pack sample (3D rendered on the left), colored according to the velocity magnitude measured in each point (right). Only particles which could be tracked for at least 20 time frames are included.}
\label{Trajectory renderings}
\end{figure}

\begin{figure}[!htb]
\centering
\includegraphics[width=\linewidth]{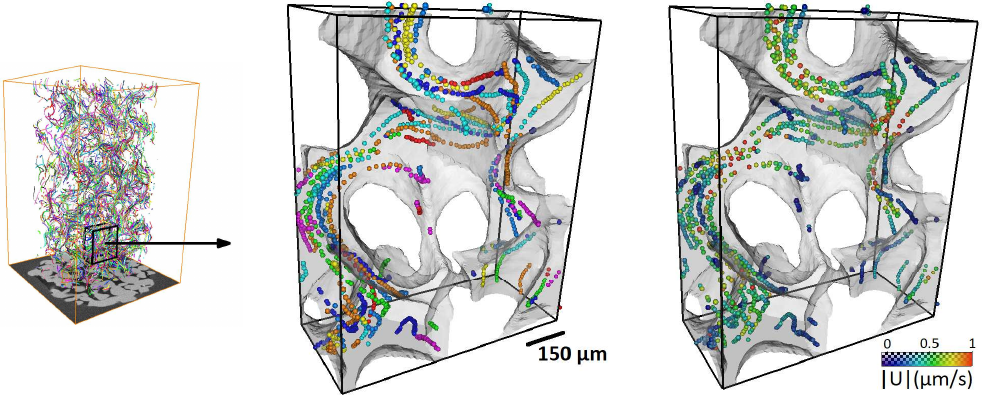}
\caption{A detailed view of the experimentally measured tracer trajectories in the porous glass filter sample, assigned a random color per individual particle (left and center) and colored according to the velocity magnitude $|U|$ measured in each point (right). Only trajectories that spanned at least 20 time frames are shown.}
\label{Trajectory renderings detail}
\end{figure}

\begin{figure}[!htb]
\centering
\includegraphics[width=0.9\linewidth]{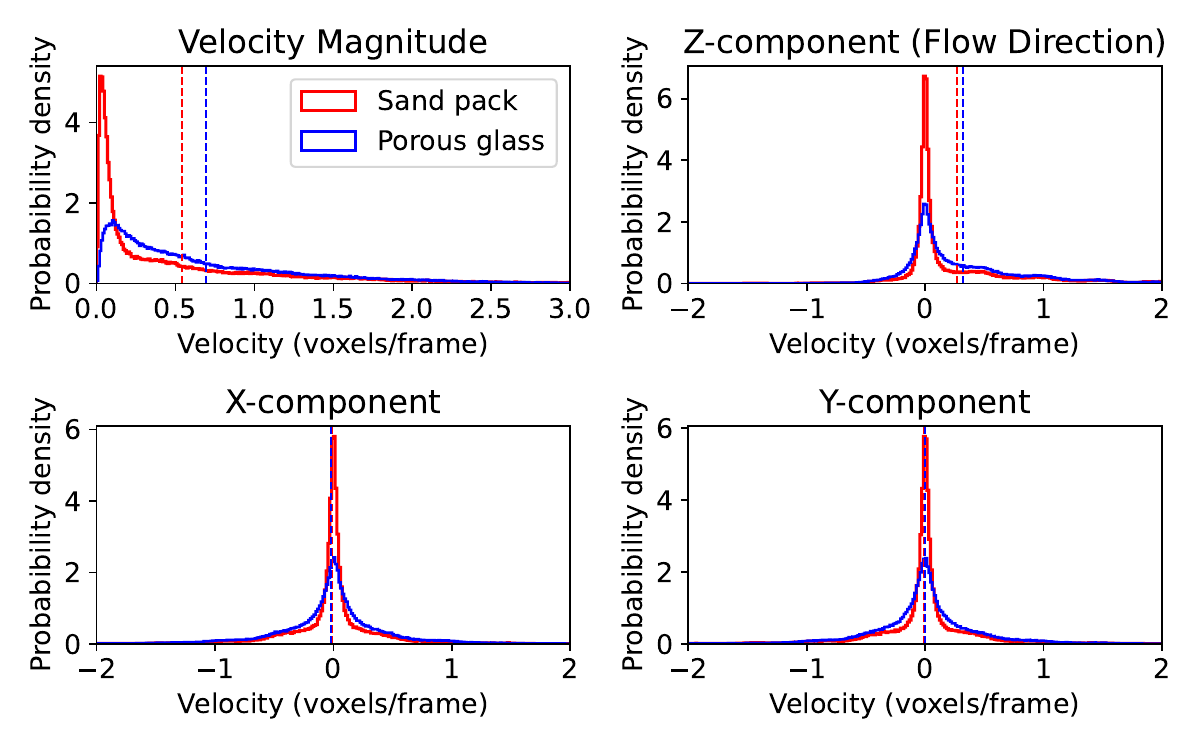}
\caption{Velocity distributions extracted from the experimentally-measured tracer trajectories for the two samples, indicated in voxels per frame (12 µm and 35 s, respectively). Mean velocities are indicated as dotted vertical lines. The mean velocity in the flow direction ( {Z}) matched well with the interstitial velocity calculated from the pump rate  {(0.54 and 0.70 voxels/frame versus 0.38 and 0.63 voxels/frame, respectively)}. The velocity components orthogonal to the flow direction are distributed symmetrically around zero, as expected.}
\label{Velocity distributions}
\end{figure}

The identified particle locations in each time frame were linked together by the nearest-neighbour algorithm, resulting in a total of 11084 (sand pack) and 50005 trajectories (porous glass), from which respectively 2490 and 4415 had the imposed minimum span of 20 time steps. 3D renderings of the filtered tracks followed tortuous paths through the pore space, as shown in Figures \ref{Trajectory renderings} and \ref{Trajectory renderings detail}. The velocities at each point in these tracks were calculated, yielding the velocity distributions in Figure \ref{Velocity distributions}. The distributions of the X- and Y-velocity components perpendicular to the global flow direction were symmetrically distributed around zero, as expected. The mean Z-components of the  {measured} velocities were resp. 0.54 and 0.70 voxel/frame  {for the two experiments} (182 and 236 nm/s), compared to the interstitial velocities of 0.38 and 0.63 voxels/frame calculated from the injection rate (Table \ref{Table fluid particle properties}). This is an encouraging match, especially since the tortuosity of the pore space was not taken into account in the interstitial velocity, meaning that the real average velocity in the pores was likely a factor between 1 and 2 larger than the interstitial velocity \citep{Fu2021}.

\begin{figure}[htb]
\centering
\includegraphics[width=1\linewidth]{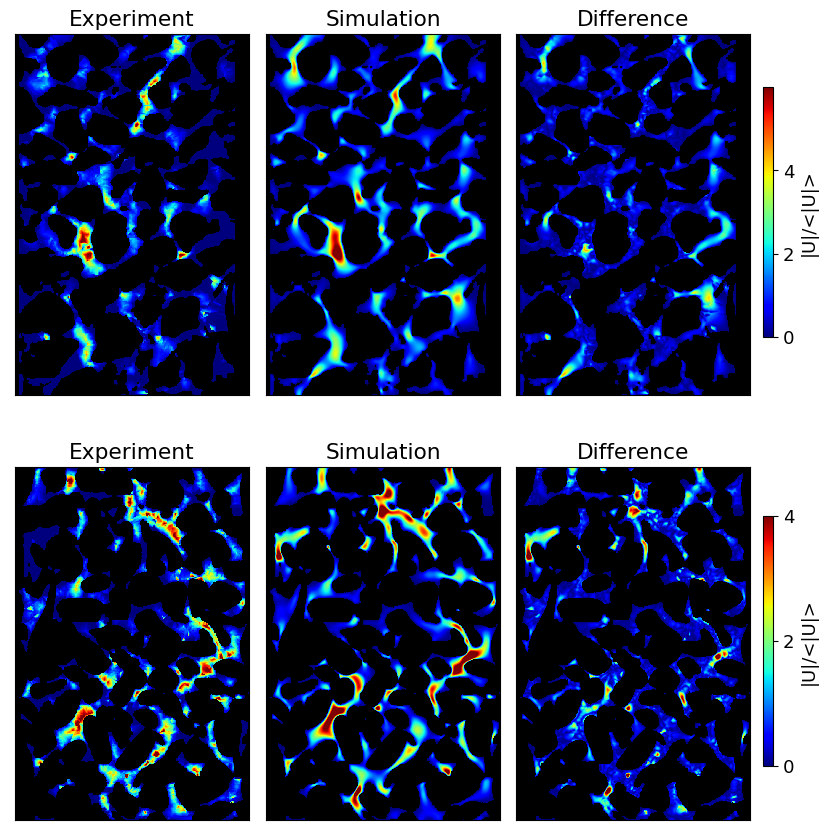}
\caption{Vertical cross-sections through the 3D, normalized velocity magnitude fields from the sand pack (top) and the porous glass experiment (bottom) matched well with computational fluid dynamics predictions. {The figures on the right shows the absolute difference between the experimental and simulated velocity fields}. Mismatches close to the sample boundary in the sand pack dataset may be due to inlet effects.}
\label{Velocity fields}
\end{figure}

Finally, the particle velocities were interpolated to find the velocity fields on a voxel grid. In Figure \ref{Velocity fields}, we compare this to CFD simulations of the velocity field using the OpenFOAM-based solver \citep{Raeini2022} mentioned in Section \ref{velocity field interpolation and CFD}. The experimental measurements and the CFD-simulations of the pore-scale velocity distributions matched well (Figure \ref{Velocity fields}). In the sand pack, mismatches close to the sample boundary may be due to inlet effects: the experimental field-of-view was selected further away from the inlet than in the porous glass filter, so that the exact inlet conditions could not be taken into account in the simulation. In the experiments, the mean distance between all measured velocity points was approximately 10 voxels, which gives an indication of the sampling density of the interpolated field. Note however that particle tracking velocimetry does not uniformly sample the velocity field, as fewer observations are made in low-velocity regions. The sampling can be refined by acquiring more time frames.

\FloatBarrier

\subsection{Validation simulation results} \label{Validation results}
Contrary to established micro-velocimetry approaches, our method used cone-beam µCT data, which may suffer from specific artifacts that could impact the detection and localization of tracer particles: the geometrical deformations at the top and bottom of the volume (\lq\lq{}cone beam artifacts\rq\rq{}), limited spatial resolution for fast image acquisition, and motion artifacts \citep{Cnudde2013, Makiharju2022}. Since the impact of these artifacts was unclear and difficult to quantify in the experimental data, we created and analyzed a digital twin of the porous glass experiment. 

\begin{figure}[htb]

\includegraphics[width=0.95\linewidth]{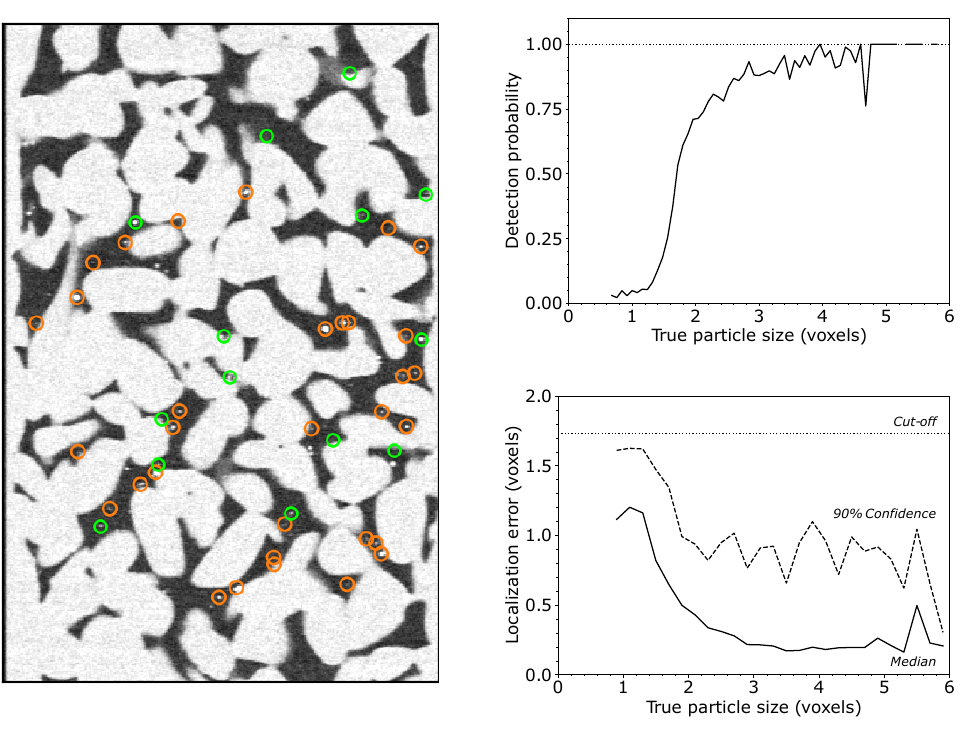}
\centering
\caption{On the left, a cross-sectional slice from the simulated validation data set, showing particle detections in the same slice in green and in the neighbouring slices in orange. On the right, the detection probability (top) and the localization error (bottom) of the ground-truth particles during frame-by-frame particle identification. In total, 48\% of the ground-truth particles were recovered, with a median localization error of 0.36 voxels (4.25 µm) from the true particle location. }
\label{Porous medium particle detection efficiency}
\end{figure}

Figure \ref{Porous medium particle detection efficiency} illustrates the detection of particles in the validation data set in function of their size. A particle was considered to be detected if there was a detection closer than $\sqrt{3}$ voxels (a voxel diagonal) from the true position. The settings used for particle detection were the same as those used in the experiments, with exception of the intensity threshold, which was slightly modified from 98 to 98.5\% to account for the fact that the experiments contained a small amount of bright particle agglomerations, which was not the case in the simulations. In total, the method detected approximately 48\% of the ground-truth particles in individual time frames, mainly dependent on the particle size (Figure \ref{Porous medium particle detection efficiency}). Particles that went undetected did not necessarily cause errors in the velocity field, but did reduce the efficiency in terms of measurement time. Approximately 6 \% of the particle detections could not be matched to a ground-truth particle. These false detections led to errors in the velocity field if they were subsequently wrongly linked into particle trajectories. Figure \ref{Porous medium particle detection efficiency} shows the localization error: the distance between the correct location of a ground-truth particle (at a time point in the middle of the acquisition) and the recovered location. Approximately 90\% of the detected particles could be localized with an {error below one voxel length} (90\% confidence error bound: 1.02 voxels). The localization error had a median of 0.36 voxels and increased significantly for smaller particles. We present these errors in units of voxels/frame because we may expect similar values in experiments with other voxel sizes or frame rates, as long the particle velocities scale accordingly (and the signal-to-noise ratio remains similar).

After linking the detected particles into trajectories and removing those shorter than 6 time steps, 33\% of the true trajectories were retrieved, meaning one particle was detected within $\sqrt{3}$ voxels of the same true trajectory for at least 6 time steps. Only 9.3\% of the detected tracks did not match a true trajectory. However, most of these \lq\lq{}false positives\rq{}\rq{} were made up of two correct parts of true tracks that were wrongly linked together, meaning they still produced at least 3 accurate velocity points for 2 incorrect ones. Less than 1\% of the detected trajectories could not be matched to 1 or 2 true trajectories. Note that a more stringent length cut-off was applied in the experiments as there were more time steps available than in the simulations, which may have resulted in more accurate linking.

\begin{figure}[!htb]
\centering
\includegraphics[width=0.9\linewidth]{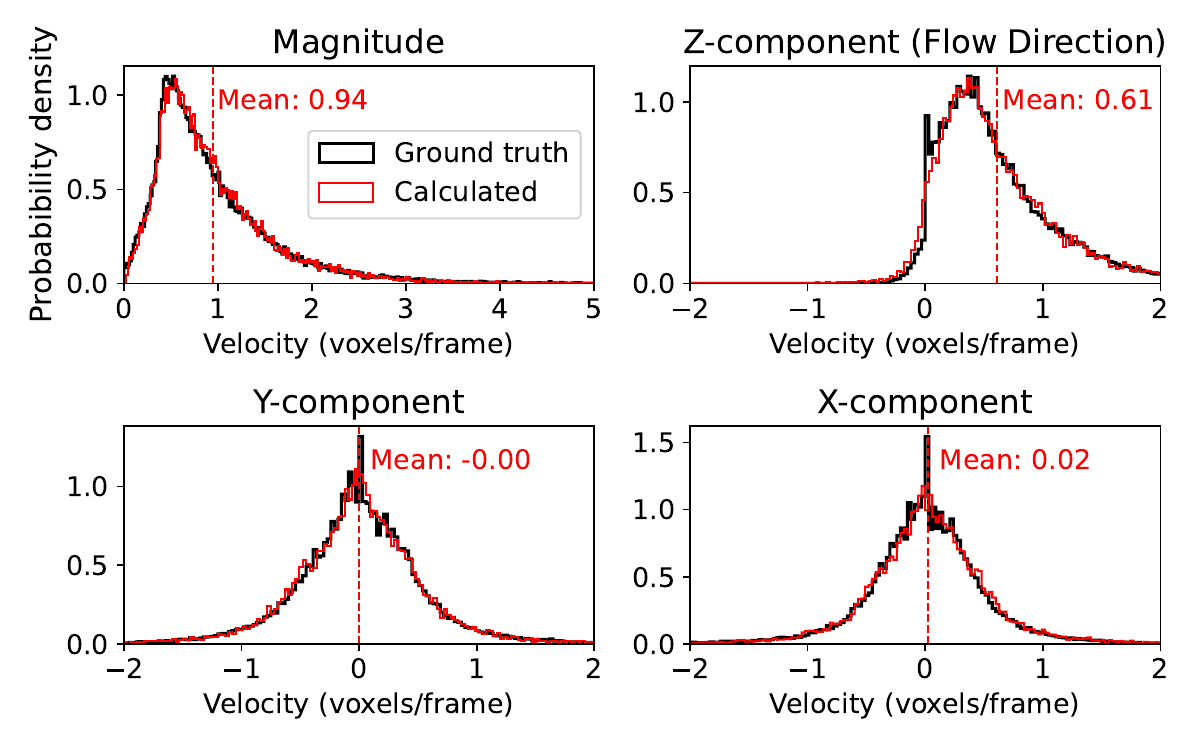}
\caption{Measured and ground-truth distributions of particle velocities in the simulated validation data set showed a close match.}
\label{Porous medium velocity distribution}
\end{figure}

\begin{figure}[!htb]
\centering
\includegraphics[width=0.9\linewidth]{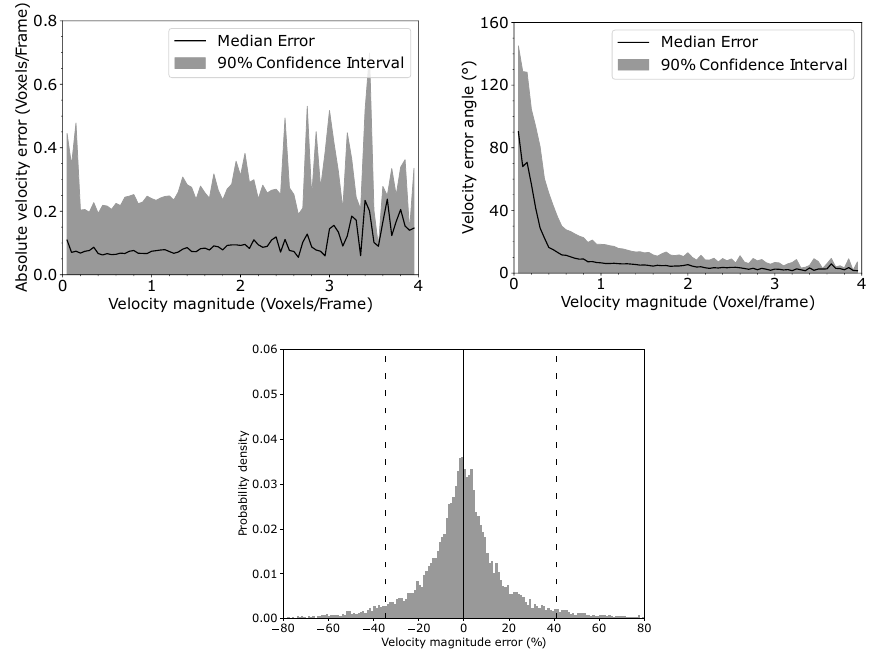}
\caption{Top left: point-by-point absolute velocity magnitude errors remained relatively constant below true magnitudes of 3 voxels per frame, after which they increased. Top right: there was a larger angle between the true and measured velocities for small particle displacements, i.e. at low velocity magnitudes (top right), due to the finite resolution of the images. Bottom: the histogram of the relative error on the velocity magntitude, with the 90\% bounds indicated as {dashed} vertical lines. }
\label{Porous medium velocity error}
\end{figure}

Figure \ref{Porous medium velocity distribution} shows that the recovered and ground-truth velocity distributions in the validation data had an excellent match. By design, these distributions were also similar to the experimental velocity magnitudes in Figure \ref{Velocity distributions}, suggesting that similar motion artifacts and trajectory linking errors can be expected. In the validation data, the median absolute error on the velocity magnitude was 0.072 voxels/frame, with a 90\% confidence error bound of 0.24 voxels/frame. This was smaller than the localization error before linking suggested, likely due to the rejection of low-quality detections during the length filtering, and potentially because some systematic errors in the localization may cancel out in the velocities. As shown in Figure \ref{Porous medium velocity error}, the absolute error remained approximately constant for true velocity magnitudes below 3 voxels/frame, after which the error started to increase. There was no systematic over- or underestimation, but the measurement did show significant scatter: the relative error bounds on the velocity magnitude (90\% confidence) were approximately $\pm40$\,\%. The directional error, i.e. the angle between the detected and the true velocity vectors, had a median value of 8.6\textdegree\ and a 90\% confidence bound of 30.0\textdegree. As shown in figure \ref{Porous medium velocity error}, the error angle was large where the velocity magnitude was below 0.5 voxels/frame, as it was difficult to accurately quantify small particle movements because of the finite resolution of the images. This was also shown by the fact that the relative error on the magnitude was larger here. However, this could be improved relatively easily, for example by skipping time steps in a particle's trajectory until it has moved more than a minimum set distance before calculating its velocity. 

\begin{figure}[!htb]
\centering
\includegraphics[width=0.9\linewidth]{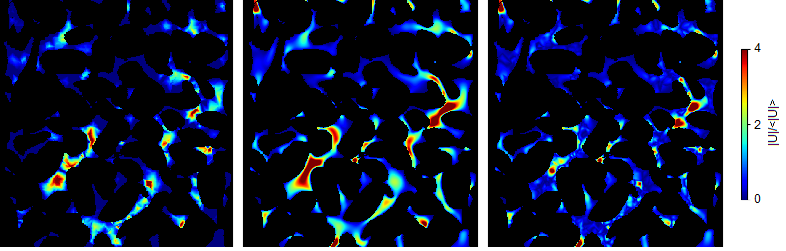}
\caption{In the simulated validation data set, the recovered (left) and the simulated (middle) velocity fields showed a comparable match as in the experiments. {The figure on the right shows the absolute difference between the experimental and simulated velocity fields.}}
\label{Porous medium velocity field}
\end{figure}

Finally, we show the interpolated velocity field for the validation data in Figure \ref{Porous medium velocity field}, for the part of the image in which particles were seeded. Visually, the match to the simulated equivalent is comparable to that in the experiments from Figure \ref{Velocity fields}, indicating the suitability of the error analyses above.

\FloatBarrier



\section{Conclusions}
In this paper, we present the first successful use of X-ray imaging to perform 3D velocimetry on flow in porous media. We presented two experiments on creeping, single-phase flow in a sandpack and in a porous sintered glass filter, in which the paths of thousands of individual tracer particles travelling through the pore space were successfully tracked. The resulting velocity field matched well with a computational fluid dynamics simulation on the same samples. The tracer particles used here were silver-coated spheres with a mean particle size around 20 µm, suspended in a viscous liquid to slow down gravitational settling. The experiments relied on continuous µCT acquisition with a voxel size of 11.8 µm and an acquisition time of 70\,s\,/\,360\textdegree\ scan, which, through an interleaved reconstruction scheme, resulted in a series of 3D images with a time (frame) interval of 35\,s. The particle trajectories were identified using a relatively straightforward nearest-neighbour algorithm based on an open-source library (TrackPy).

The results were validated with the help of a digital twin of the porous glass experiment, created by numerically simulating the µCT imaging of particles as they move through the pores. Due to the small particle size compared to the voxel size, approximately 50\% of the simulated particles could be detected in each image. However, particles that were large enough to be detected could be localized with an accuracy below the voxel size in 90\% of the cases. From the recovered particle trajectories, we were able to measure velocity magnitudes up to approximately 4 voxels/frame (0.69 µm/s) with an error below 0.24 voxels/frame (0.04 µm/s; 90\% confidence). The recovered velocity vectors were inaccurate for small velocities ($<$ 0.5 voxels/frame) as the particle displacements per individual time frame were then too small compared to the resolution - an issue which may be resolved by better post-processing. These validation results are expected to hold general validity towards these and other similar experiments. The main source of errors that could not be taken into account in the validation were mechanical and electronic inaccuracies of the scanner. These were deemed secondary to photon counting noise and motion artifacts for the fast imaging with relatively large voxel sizes presented here, but may still have caused the errors in the experimental data to be larger than in the validation. 

Our work proves the feasibility of µCT-based particle velocimetry in complex geometries, and suggests that there is a large potential for further development and application of this method. While our measurements were limited to low flow rates, highly viscous liquids and samples with large pores, these were not hard limitations. At synchrotron beam lines, imaging at voxel sizes up to approximately 4 times smaller with acquisition times 100 times faster have become routinely possible \citep{Spurin2021}, meaning velocities of up to 2 orders of magnitude larger than in this work could be measured. In both laboratory-based and synchrotron µCT, the imaging can be sped up further by advanced reconstruction methods using e.g. prior knowledge on the process \citep{Myers2011, VanEyndhoven2015} and motion-compensation \citep{DeSchryver2018}. Furthermore, the higher spatial resolutions that can be achieved using these approaches would facilitate the use of smaller and lighter tracer particles, thereby also easing the limitations on the viscosity of the liquid and on the sample's pore size. The particle detection and linking scheme applied in this paper can also still be improved using more sophisticated methods \citep{Chenouard2014}. There is ample opportunity to apply all of the above concepts to µCT-based velocimetry. The resulting methods could bring forth a turning point in the study of fluid dynamics in complex, microscopic geometries; ranging from porous materials to (bio-)medical applications and industrial fluid flows. 

\section*{Appendix}\label{Appendix}
\begin{figure}[!htb]
\centering
\includegraphics[width=0.45\linewidth]{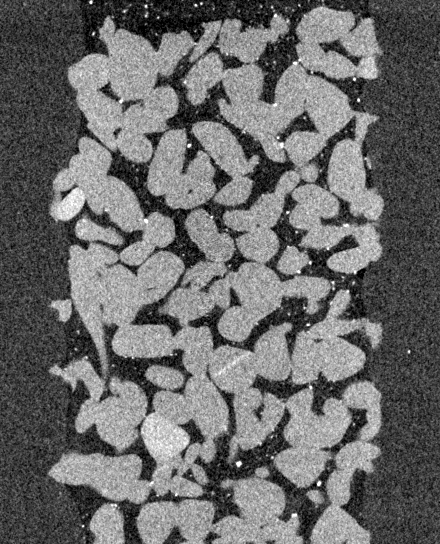}
\caption{A cross-sectional view of the reconstructed time frames in the porous glass experiment, showing brightly colored tracer particles (multimedia view).}
\label{cross-section grey value video}
\end{figure}

\begin{figure}[!htb]
\centering
  \begin{subfigure}{0.4\linewidth}
  \includegraphics[width=\linewidth]{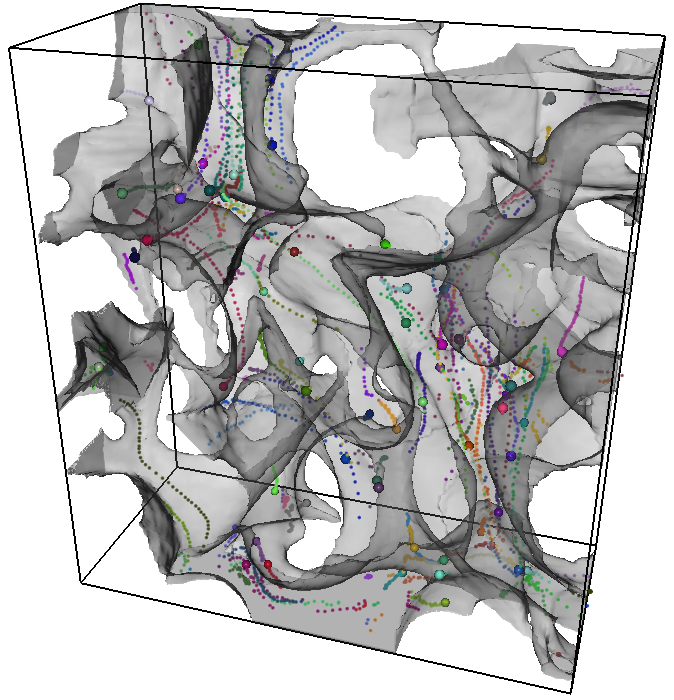}
  \caption{}
  \end{subfigure}
  \begin{subfigure}{0.4\linewidth}
    \includegraphics[width=\linewidth]{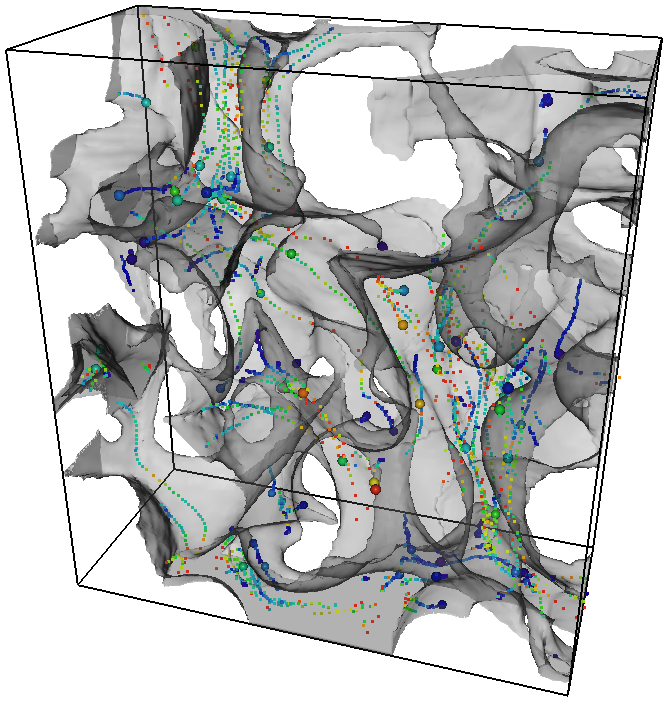}
  \caption{}
  \end{subfigure}

\caption{Particles moving along their tracks in a detail of the porous glass dataset. In figure a (multimedia view), each individual particle is assigned a random color. In figure b (multimedia view), the color scale reflects the velocity magnitude in each point.}
\label{detail particle moving video}
\end{figure}

To show how tracer particles moved smoothly through the pores during the velocimetry experiments, a video of the central vertical cross-sectional slice (parallel to the flow direction) in each time frame of the porous glass experiment  is provided in Figure \ref{cross-section grey value video} (multimedia view). The identification of trajectories on this dataset led to the videos shown in Figure \ref{detail particle moving video}a (multimedia view) and \ref{detail particle moving video}b (multimedia view), showing a detail from the full dataset, where trajectories were colored by particle or by local velocity magnitude.

\FloatBarrier

\begin{acknowledgments}
Dr. Inka Meyer (Ghent University) is thanked for her help with measuring the tracer particle size distribution. Steffen Berg and co-workers at Shell are thanked for inspiring discussions around velocimetry in porous media. Tom Bultreys holds a senior postdoctoral fellowship from the Research Foundation Flanders (FWO) under grant 12X0922N. This research was also partially funded under the Strategic Basic Research Program MoCCha-CT (S003418N) and the Junior Research Project program (3G036518) of the Research Foundation - Flanders. 

\end{acknowledgments}

\section*{Data Availability Statement}
{The data that support the findings of this study are freely available} from Zenodo. This includes the full experimental data sets containing the 3D time frames and the resulting particle trajectory data and velocity fields:
\begin{itemize}
    \item Sand pack experiment: http://doi.org/10.5281/zenodo.6010425
    \item Porous glass experiment: http://doi.org/10.5281/zenodo.6010490
    \item Validation simulations: http://doi.org/10.5281/zenodo.6010914
\end{itemize}

\bibliography{mybibliography}

\begin{thebibliography}{59}%
\makeatletter
\providecommand \@ifxundefined [1]{%
 \@ifx{#1\undefined}
}%
\providecommand \@ifnum [1]{%
 \ifnum #1\expandafter \@firstoftwo
 \else \expandafter \@secondoftwo
 \fi
}%
\providecommand \@ifx [1]{%
 \ifx #1\expandafter \@firstoftwo
 \else \expandafter \@secondoftwo
 \fi
}%
\providecommand \natexlab [1]{#1}%
\providecommand \enquote  [1]{``#1''}%
\providecommand \bibnamefont  [1]{#1}%
\providecommand \bibfnamefont [1]{#1}%
\providecommand \citenamefont [1]{#1}%
\providecommand \href@noop [0]{\@secondoftwo}%
\providecommand \href [0]{\begingroup \@sanitize@url \@href}%
\providecommand \@href[1]{\@@startlink{#1}\@@href}%
\providecommand \@@href[1]{\endgroup#1\@@endlink}%
\providecommand \@sanitize@url [0]{\catcode `\\12\catcode `\$12\catcode
  `\&12\catcode `\#12\catcode `\^12\catcode `\_12\catcode `\%12\relax}%
\providecommand \@@startlink[1]{}%
\providecommand \@@endlink[0]{}%
\providecommand \url  [0]{\begingroup\@sanitize@url \@url }%
\providecommand \@url [1]{\endgroup\@href {#1}{\urlprefix }}%
\providecommand \urlprefix  [0]{URL }%
\providecommand \Eprint [0]{\href }%
\providecommand \doibase [0]{http://dx.doi.org/}%
\providecommand \selectlanguage [0]{\@gobble}%
\providecommand \bibinfo  [0]{\@secondoftwo}%
\providecommand \bibfield  [0]{\@secondoftwo}%
\providecommand \translation [1]{[#1]}%
\providecommand \BibitemOpen [0]{}%
\providecommand \bibitemStop [0]{}%
\providecommand \bibitemNoStop [0]{.\EOS\space}%
\providecommand \EOS [0]{\spacefactor3000\relax}%
\providecommand \BibitemShut  [1]{\csname bibitem#1\endcsname}%
\let\auto@bib@innerbib\@empty
\bibitem [{\citenamefont {An}\ \emph {et~al.}(2022)\citenamefont {An},
  \citenamefont {Sahimi}, \citenamefont {Shende}, \citenamefont {Babaei},\ and\
  \citenamefont {Niasar}}]{An2022}%
  \BibitemOpen
  \bibfield  {author} {\bibinfo {author} {\bibnamefont {An}, \bibfnamefont
  {S.}}, \bibinfo {author} {\bibnamefont {Sahimi}, \bibfnamefont {M.}},
  \bibinfo {author} {\bibnamefont {Shende}, \bibfnamefont {T.}}, \bibinfo
  {author} {\bibnamefont {Babaei}, \bibfnamefont {M.}}, \ and\ \bibinfo
  {author} {\bibnamefont {Niasar}, \bibfnamefont {V.}},\ }\bibfield  {title}
  {\enquote {\bibinfo {title} {Enhanced thermal fingering in a shear-thinning
  fluid flow through porous media: Dynamic pore network modeling},}\ }\href
  {\doibase 10.1063/5.0080375} {\bibfield  {journal} {\bibinfo  {journal}
  {Physics of Fluids}\ }\textbf {\bibinfo {volume} {34}},\ \bibinfo {pages}
  {023105} (\bibinfo {year} {2022})}\BibitemShut {NoStop}%
\bibitem [{\citenamefont {Baker}\ \emph {et~al.}(2018)\citenamefont {Baker},
  \citenamefont {Guillard}, \citenamefont {Marks},\ and\ \citenamefont
  {Einav}}]{Baker2018}%
  \BibitemOpen
  \bibfield  {author} {\bibinfo {author} {\bibnamefont {Baker}, \bibfnamefont
  {J.}}, \bibinfo {author} {\bibnamefont {Guillard}, \bibfnamefont {F.}},
  \bibinfo {author} {\bibnamefont {Marks}, \bibfnamefont {B.}}, \ and\ \bibinfo
  {author} {\bibnamefont {Einav}, \bibfnamefont {I.}},\ }\bibfield  {title}
  {\enquote {\bibinfo {title} {X-ray rheography uncovers planar granular flows
  despite non-planar walls},}\ }\href {\doibase 10.1038/s41467-018-07628-6}
  {\bibfield  {journal} {\bibinfo  {journal} {Nature Communications}\ }\textbf
  {\bibinfo {volume} {9}},\ \bibinfo {pages} {1--9} (\bibinfo {year}
  {2018})}\BibitemShut {NoStop}%
\bibitem [{\citenamefont {Berg}\ \emph {et~al.}(2013)\citenamefont {Berg},
  \citenamefont {Ott}, \citenamefont {a~Klapp}, \citenamefont {Schwing},
  \citenamefont {Neiteler}, \citenamefont {Brussee}, \citenamefont {Makurat},
  \citenamefont {Leu}, \citenamefont {Enzmann}, \citenamefont {Schwarz},
  \citenamefont {Kersten}, \citenamefont {Irvine},\ and\ \citenamefont
  {Stampanoni}}]{Berg2013}%
  \BibitemOpen
  \bibfield  {author} {\bibinfo {author} {\bibnamefont {Berg}, \bibfnamefont
  {S.}}, \bibinfo {author} {\bibnamefont {Ott}, \bibfnamefont {H.}}, \bibinfo
  {author} {\bibnamefont {a~Klapp}, \bibfnamefont {S.}}, \bibinfo {author}
  {\bibnamefont {Schwing}, \bibfnamefont {A.}}, \bibinfo {author} {\bibnamefont
  {Neiteler}, \bibfnamefont {R.}}, \bibinfo {author} {\bibnamefont {Brussee},
  \bibfnamefont {N.}}, \bibinfo {author} {\bibnamefont {Makurat}, \bibfnamefont
  {A.}}, \bibinfo {author} {\bibnamefont {Leu}, \bibfnamefont {L.}}, \bibinfo
  {author} {\bibnamefont {Enzmann}, \bibfnamefont {F.}}, \bibinfo {author}
  {\bibnamefont {Schwarz}, \bibfnamefont {J.-O.}}, \bibinfo {author}
  {\bibnamefont {Kersten}, \bibfnamefont {M.}}, \bibinfo {author} {\bibnamefont
  {Irvine}, \bibfnamefont {S.}}, \ and\ \bibinfo {author} {\bibnamefont
  {Stampanoni}, \bibfnamefont {M.}},\ }\bibfield  {title} {\enquote {\bibinfo
  {title} {Real-time 3d imaging of haines jumps in porous media flow},}\ }\href
  {\doibase 10.1073/pnas.1221373110} {\bibfield  {journal} {\bibinfo  {journal}
  {Proceedings of the National Academy of Sciences}\ }\textbf {\bibinfo
  {volume} {110}},\ \bibinfo {pages} {3755--3759} (\bibinfo {year}
  {2013})}\BibitemShut {NoStop}%
\bibitem [{\citenamefont {Blunt}(2017)}]{Blunt2017}%
  \BibitemOpen
  \bibfield  {author} {\bibinfo {author} {\bibnamefont {Blunt}, \bibfnamefont
  {M.~J.}},\ }\href@noop {} {\emph {\bibinfo {title} {Multiphase Flow in
  Permeable Media: A Pore-Scale Perspective}}}\ (\bibinfo  {publisher}
  {Cambridge University Press},\ \bibinfo {year} {2017})\BibitemShut {NoStop}%
\bibitem [{\citenamefont {Bui}\ \emph {et~al.}(2018)\citenamefont {Bui},
  \citenamefont {Adjiman}, \citenamefont {Bardow}, \citenamefont {Anthony},
  \citenamefont {Boston}, \citenamefont {Brown}, \citenamefont {Fennell},
  \citenamefont {Fuss}, \citenamefont {Galindo}, \citenamefont {Hackett},
  \citenamefont {Hallett}, \citenamefont {Herzog}, \citenamefont {Jackson},
  \citenamefont {Kemper}, \citenamefont {Krevor}, \citenamefont {Maitland},
  \citenamefont {Matuszewski}, \citenamefont {Metcalfe}, \citenamefont {Petit},
  \citenamefont {Puxty}, \citenamefont {Reimer}, \citenamefont {Reiner},
  \citenamefont {Rubin}, \citenamefont {Scott}, \citenamefont {Shah},
  \citenamefont {Smit}, \citenamefont {Trusler}, \citenamefont {Webley},
  \citenamefont {Wilcox},\ and\ \citenamefont {Dowell}}]{Bui2018}%
  \BibitemOpen
  \bibfield  {author} {\bibinfo {author} {\bibnamefont {Bui}, \bibfnamefont
  {M.}}, \bibinfo {author} {\bibnamefont {Adjiman}, \bibfnamefont {C.~S.}},
  \bibinfo {author} {\bibnamefont {Bardow}, \bibfnamefont {A.}}, \bibinfo
  {author} {\bibnamefont {Anthony}, \bibfnamefont {E.~J.}}, \bibinfo {author}
  {\bibnamefont {Boston}, \bibfnamefont {A.}}, \bibinfo {author} {\bibnamefont
  {Brown}, \bibfnamefont {S.}}, \bibinfo {author} {\bibnamefont {Fennell},
  \bibfnamefont {P.~S.}}, \bibinfo {author} {\bibnamefont {Fuss}, \bibfnamefont
  {S.}}, \bibinfo {author} {\bibnamefont {Galindo}, \bibfnamefont {A.}},
  \bibinfo {author} {\bibnamefont {Hackett}, \bibfnamefont {L.~A.}}, \bibinfo
  {author} {\bibnamefont {Hallett}, \bibfnamefont {J.~P.}}, \bibinfo {author}
  {\bibnamefont {Herzog}, \bibfnamefont {H.~J.}}, \bibinfo {author}
  {\bibnamefont {Jackson}, \bibfnamefont {G.}}, \bibinfo {author} {\bibnamefont
  {Kemper}, \bibfnamefont {J.}}, \bibinfo {author} {\bibnamefont {Krevor},
  \bibfnamefont {S.}}, \bibinfo {author} {\bibnamefont {Maitland},
  \bibfnamefont {G.~C.}}, \bibinfo {author} {\bibnamefont {Matuszewski},
  \bibfnamefont {M.}}, \bibinfo {author} {\bibnamefont {Metcalfe},
  \bibfnamefont {I.~S.}}, \bibinfo {author} {\bibnamefont {Petit},
  \bibfnamefont {C.}}, \bibinfo {author} {\bibnamefont {Puxty}, \bibfnamefont
  {G.}}, \bibinfo {author} {\bibnamefont {Reimer}, \bibfnamefont {J.}},
  \bibinfo {author} {\bibnamefont {Reiner}, \bibfnamefont {D.~M.}}, \bibinfo
  {author} {\bibnamefont {Rubin}, \bibfnamefont {E.~S.}}, \bibinfo {author}
  {\bibnamefont {Scott}, \bibfnamefont {S.~A.}}, \bibinfo {author}
  {\bibnamefont {Shah}, \bibfnamefont {N.}}, \bibinfo {author} {\bibnamefont
  {Smit}, \bibfnamefont {B.}}, \bibinfo {author} {\bibnamefont {Trusler},
  \bibfnamefont {J.~P.}}, \bibinfo {author} {\bibnamefont {Webley},
  \bibfnamefont {P.}}, \bibinfo {author} {\bibnamefont {Wilcox}, \bibfnamefont
  {J.}}, \ and\ \bibinfo {author} {\bibnamefont {Dowell}, \bibfnamefont
  {N.~M.}},\ }\bibfield  {title} {\enquote {\bibinfo {title} {Carbon capture
  and storage (ccs): The way forward},}\ }\href {\doibase 10.1039/c7ee02342a}
  {\bibfield  {journal} {\bibinfo  {journal} {Energy and Environmental
  Science}\ }\textbf {\bibinfo {volume} {11}},\ \bibinfo {pages} {1062--1176}
  (\bibinfo {year} {2018})}\BibitemShut {NoStop}%
\bibitem [{\citenamefont {Bultreys}\ \emph {et~al.}(2016)\citenamefont
  {Bultreys}, \citenamefont {Boone}, \citenamefont {Boone}, \citenamefont
  {Schryver}, \citenamefont {Masschaele}, \citenamefont {Hoorebeke},\ and\
  \citenamefont {Cnudde}}]{Bultreys2015f}%
  \BibitemOpen
  \bibfield  {author} {\bibinfo {author} {\bibnamefont {Bultreys},
  \bibfnamefont {T.}}, \bibinfo {author} {\bibnamefont {Boone}, \bibfnamefont
  {M.~A.}}, \bibinfo {author} {\bibnamefont {Boone}, \bibfnamefont {M.~N.}},
  \bibinfo {author} {\bibnamefont {Schryver}, \bibfnamefont {T.~D.}}, \bibinfo
  {author} {\bibnamefont {Masschaele}, \bibfnamefont {B.}}, \bibinfo {author}
  {\bibnamefont {Hoorebeke}, \bibfnamefont {L.~V.}}, \ and\ \bibinfo {author}
  {\bibnamefont {Cnudde}, \bibfnamefont {V.}},\ }\bibfield  {title} {\enquote
  {\bibinfo {title} {Fast laboratory-based micro-computed tomography for
  pore-scale research: Illustrative experiments and perspectives on the
  future},}\ }\href {\doibase 10.1016/j.advwatres.2015.05.012} {\bibfield
  {journal} {\bibinfo  {journal} {Advances in Water Resources}\ }\textbf
  {\bibinfo {volume} {95}},\ \bibinfo {pages} {341--351} (\bibinfo {year}
  {2016})}\BibitemShut {NoStop}%
\bibitem [{\citenamefont {Chenouard}\ \emph {et~al.}(2014)\citenamefont
  {Chenouard}, \citenamefont {Smal}, \citenamefont {de~Chaumont}, \citenamefont
  {Maška}, \citenamefont {Sbalzarini}, \citenamefont {Gong}, \citenamefont
  {Cardinale}, \citenamefont {Carthel}, \citenamefont {Coraluppi},
  \citenamefont {Winter}, \citenamefont {Cohen}, \citenamefont {Godinez},
  \citenamefont {Rohr}, \citenamefont {Kalaidzidis}, \citenamefont {Liang},
  \citenamefont {Duncan}, \citenamefont {Shen}, \citenamefont {Xu},
  \citenamefont {Magnusson}, \citenamefont {Jaldén}, \citenamefont {Blau},
  \citenamefont {Paul-Gilloteaux}, \citenamefont {Roudot}, \citenamefont
  {Kervrann}, \citenamefont {Waharte}, \citenamefont {Tinevez}, \citenamefont
  {Shorte}, \citenamefont {Willemse}, \citenamefont {Celler}, \citenamefont
  {van Wezel}, \citenamefont {Dan}, \citenamefont {Tsai}, \citenamefont
  {de~Solórzano}, \citenamefont {Olivo-Marin},\ and\ \citenamefont
  {Meijering}}]{Chenouard2014}%
  \BibitemOpen
  \bibfield  {author} {\bibinfo {author} {\bibnamefont {Chenouard},
  \bibfnamefont {N.}}, \bibinfo {author} {\bibnamefont {Smal}, \bibfnamefont
  {I.}}, \bibinfo {author} {\bibnamefont {de~Chaumont}, \bibfnamefont {F.}},
  \bibinfo {author} {\bibnamefont {Maška}, \bibfnamefont {M.}}, \bibinfo
  {author} {\bibnamefont {Sbalzarini}, \bibfnamefont {I.~F.}}, \bibinfo
  {author} {\bibnamefont {Gong}, \bibfnamefont {Y.}}, \bibinfo {author}
  {\bibnamefont {Cardinale}, \bibfnamefont {J.}}, \bibinfo {author}
  {\bibnamefont {Carthel}, \bibfnamefont {C.}}, \bibinfo {author} {\bibnamefont
  {Coraluppi}, \bibfnamefont {S.}}, \bibinfo {author} {\bibnamefont {Winter},
  \bibfnamefont {M.}}, \bibinfo {author} {\bibnamefont {Cohen}, \bibfnamefont
  {A.~R.}}, \bibinfo {author} {\bibnamefont {Godinez}, \bibfnamefont {W.~J.}},
  \bibinfo {author} {\bibnamefont {Rohr}, \bibfnamefont {K.}}, \bibinfo
  {author} {\bibnamefont {Kalaidzidis}, \bibfnamefont {Y.}}, \bibinfo {author}
  {\bibnamefont {Liang}, \bibfnamefont {L.}}, \bibinfo {author} {\bibnamefont
  {Duncan}, \bibfnamefont {J.}}, \bibinfo {author} {\bibnamefont {Shen},
  \bibfnamefont {H.}}, \bibinfo {author} {\bibnamefont {Xu}, \bibfnamefont
  {Y.}}, \bibinfo {author} {\bibnamefont {Magnusson}, \bibfnamefont
  {K.~E.~G.}}, \bibinfo {author} {\bibnamefont {Jaldén}, \bibfnamefont {J.}},
  \bibinfo {author} {\bibnamefont {Blau}, \bibfnamefont {H.~M.}}, \bibinfo
  {author} {\bibnamefont {Paul-Gilloteaux}, \bibfnamefont {P.}}, \bibinfo
  {author} {\bibnamefont {Roudot}, \bibfnamefont {P.}}, \bibinfo {author}
  {\bibnamefont {Kervrann}, \bibfnamefont {C.}}, \bibinfo {author}
  {\bibnamefont {Waharte}, \bibfnamefont {F.}}, \bibinfo {author} {\bibnamefont
  {Tinevez}, \bibfnamefont {J.-Y.}}, \bibinfo {author} {\bibnamefont {Shorte},
  \bibfnamefont {S.~L.}}, \bibinfo {author} {\bibnamefont {Willemse},
  \bibfnamefont {J.}}, \bibinfo {author} {\bibnamefont {Celler}, \bibfnamefont
  {K.}}, \bibinfo {author} {\bibnamefont {van Wezel}, \bibfnamefont {G.~P.}},
  \bibinfo {author} {\bibnamefont {Dan}, \bibfnamefont {H.-W.}}, \bibinfo
  {author} {\bibnamefont {Tsai}, \bibfnamefont {Y.-S.}}, \bibinfo {author}
  {\bibnamefont {de~Solórzano}, \bibfnamefont {C.~O.}}, \bibinfo {author}
  {\bibnamefont {Olivo-Marin}, \bibfnamefont {J.-C.}}, \ and\ \bibinfo {author}
  {\bibnamefont {Meijering}, \bibfnamefont {E.}},\ }\bibfield  {title}
  {\enquote {\bibinfo {title} {Objective comparison of particle tracking
  methods.}}\ }\href {\doibase 10.1038/nmeth.2808} {\bibfield  {journal}
  {\bibinfo  {journal} {Nature methods}\ }\textbf {\bibinfo {volume} {11}},\
  \bibinfo {pages} {281--289} (\bibinfo {year} {2014})}\BibitemShut {NoStop}%
\bibitem [{\citenamefont {Cnudde}\ and\ \citenamefont
  {Boone}(2013)}]{Cnudde2013}%
  \BibitemOpen
  \bibfield  {author} {\bibinfo {author} {\bibnamefont {Cnudde}, \bibfnamefont
  {V.}}\ and\ \bibinfo {author} {\bibnamefont {Boone}, \bibfnamefont {M.~N.}},\
  }\bibfield  {title} {\enquote {\bibinfo {title} {High-resolution x-ray
  computed tomography in geosciences: A review of the current technology and
  applications},}\ }\href {\doibase 10.1016/j.earscirev.2013.04.003} {\bibfield
   {journal} {\bibinfo  {journal} {Earth-Science Reviews}\ }\textbf {\bibinfo
  {volume} {123}},\ \bibinfo {pages} {1--17} (\bibinfo {year}
  {2013})}\BibitemShut {NoStop}%
\bibitem [{\citenamefont {Crocker}\ and\ \citenamefont
  {Grier}(1996)}]{Crocker1996}%
  \BibitemOpen
  \bibfield  {author} {\bibinfo {author} {\bibnamefont {Crocker}, \bibfnamefont
  {J.~C.}}\ and\ \bibinfo {author} {\bibnamefont {Grier}, \bibfnamefont
  {D.~G.}},\ }\bibfield  {title} {\enquote {\bibinfo {title} {Methods of
  digital video microscopy for colloidal studies},}\ }\href {\doibase
  10.1006/jcis.1996.0217} {\bibfield  {journal} {\bibinfo  {journal} {Journal
  of Colloid and Interface Science}\ }\textbf {\bibinfo {volume} {179}},\
  \bibinfo {pages} {298--310} (\bibinfo {year} {1996})}\BibitemShut {NoStop}%
\bibitem [{\citenamefont {Datta}\ \emph {et~al.}(2013)\citenamefont {Datta},
  \citenamefont {Chiang}, \citenamefont {Ramakrishnan},\ and\ \citenamefont
  {Weitz}}]{Datta2013}%
  \BibitemOpen
  \bibfield  {author} {\bibinfo {author} {\bibnamefont {Datta}, \bibfnamefont
  {S.~S.}}, \bibinfo {author} {\bibnamefont {Chiang}, \bibfnamefont {H.}},
  \bibinfo {author} {\bibnamefont {Ramakrishnan}, \bibfnamefont {T.~S.}}, \
  and\ \bibinfo {author} {\bibnamefont {Weitz}, \bibfnamefont {D.~A.}},\
  }\bibfield  {title} {\enquote {\bibinfo {title} {Spatial fluctuations of
  fluid velocities in flow through a three-dimensional porous medium},}\ }\href
  {\doibase 10.1103/PhysRevLett.111.064501} {\bibfield  {journal} {\bibinfo
  {journal} {Physical Review Letters}\ }\textbf {\bibinfo {volume} {111}},\
  \bibinfo {pages} {1--5} (\bibinfo {year} {2013})}\BibitemShut {NoStop}%
\bibitem [{\citenamefont {Datta}, \citenamefont {Dupin},\ and\ \citenamefont
  {Weitz}(2014)}]{Datta2014}%
  \BibitemOpen
  \bibfield  {author} {\bibinfo {author} {\bibnamefont {Datta}, \bibfnamefont
  {S.~S.}}, \bibinfo {author} {\bibnamefont {Dupin}, \bibfnamefont {J.~B.}}, \
  and\ \bibinfo {author} {\bibnamefont {Weitz}, \bibfnamefont {D.~A.}},\
  }\bibfield  {title} {\enquote {\bibinfo {title} {Fluid breakup during
  simultaneous two-phase flow through a three-dimensional porous medium},}\
  }\href {\doibase 10.1063/1.4884955} {\bibfield  {journal} {\bibinfo
  {journal} {Physics of Fluids}\ }\textbf {\bibinfo {volume} {26}},\ \bibinfo
  {pages} {1--13} (\bibinfo {year} {2014})}\BibitemShut {NoStop}%
\bibitem [{\citenamefont {Datta}, \citenamefont {Ramakrishnan},\ and\
  \citenamefont {Weitz}(2014)}]{Datta2014a}%
  \BibitemOpen
  \bibfield  {author} {\bibinfo {author} {\bibnamefont {Datta}, \bibfnamefont
  {S.~S.}}, \bibinfo {author} {\bibnamefont {Ramakrishnan}, \bibfnamefont
  {T.~S.}}, \ and\ \bibinfo {author} {\bibnamefont {Weitz}, \bibfnamefont
  {D.~A.}},\ }\bibfield  {title} {\enquote {\bibinfo {title} {Mobilization of a
  trapped non-wetting fluid from a three-dimensional porous medium},}\ }\href
  {\doibase 10.1063/1.4866641} {\bibfield  {journal} {\bibinfo  {journal}
  {Physics of Fluids}\ }\textbf {\bibinfo {volume} {26}},\ \bibinfo {pages}
  {1--22} (\bibinfo {year} {2014})}\BibitemShut {NoStop}%
\bibitem [{\citenamefont {Dierick}\ \emph {et~al.}(2014)\citenamefont
  {Dierick}, \citenamefont {Loo}, \citenamefont {Masschaele}, \citenamefont
  {den Bulcke}, \citenamefont {Acker}, \citenamefont {Cnudde},\ and\
  \citenamefont {Hoorebeke}}]{Dierick2014}%
  \BibitemOpen
  \bibfield  {author} {\bibinfo {author} {\bibnamefont {Dierick}, \bibfnamefont
  {M.}}, \bibinfo {author} {\bibnamefont {Loo}, \bibfnamefont {D.~V.}},
  \bibinfo {author} {\bibnamefont {Masschaele}, \bibfnamefont {B.}}, \bibinfo
  {author} {\bibnamefont {den Bulcke}, \bibfnamefont {J.~V.}}, \bibinfo
  {author} {\bibnamefont {Acker}, \bibfnamefont {J.~V.}}, \bibinfo {author}
  {\bibnamefont {Cnudde}, \bibfnamefont {V.}}, \ and\ \bibinfo {author}
  {\bibnamefont {Hoorebeke}, \bibfnamefont {L.~V.}},\ }\bibfield  {title}
  {\enquote {\bibinfo {title} {Recent micro-ct scanner developments at ugct},}\
  }\href {\doibase 10.1016/j.nimb.2013.10.051} {\bibfield  {journal} {\bibinfo
  {journal} {Nuclear Instruments and Methods in Physics Research Section B:
  Beam Interactions with Materials and Atoms}\ }\textbf {\bibinfo {volume}
  {324}},\ \bibinfo {pages} {35--40} (\bibinfo {year} {2014})}\BibitemShut
  {NoStop}%
\bibitem [{\citenamefont {Discetti}\ and\ \citenamefont
  {Coletti}(2018)}]{Discetti2018}%
  \BibitemOpen
  \bibfield  {author} {\bibinfo {author} {\bibnamefont {Discetti},
  \bibfnamefont {S.}}\ and\ \bibinfo {author} {\bibnamefont {Coletti},
  \bibfnamefont {F.}},\ }\bibfield  {title} {\enquote {\bibinfo {title}
  {Volumetric velocimetry for fluid flows},}\ }\href {\doibase
  10.1088/1361-6501/aaa571} {\bibfield  {journal} {\bibinfo  {journal}
  {Measurement Science and Technology}\ }\textbf {\bibinfo {volume} {29}},\
  \bibinfo {pages} {042001} (\bibinfo {year} {2018})}\BibitemShut {NoStop}%
\bibitem [{\citenamefont {Dubsky}\ \emph {et~al.}(2012)\citenamefont {Dubsky},
  \citenamefont {Jamison}, \citenamefont {Higgins}, \citenamefont {Siu},
  \citenamefont {Hourigan},\ and\ \citenamefont {Fouras}}]{Dubsky2012}%
  \BibitemOpen
  \bibfield  {author} {\bibinfo {author} {\bibnamefont {Dubsky}, \bibfnamefont
  {S.}}, \bibinfo {author} {\bibnamefont {Jamison}, \bibfnamefont {R.~A.}},
  \bibinfo {author} {\bibnamefont {Higgins}, \bibfnamefont {S.~P.~A.}},
  \bibinfo {author} {\bibnamefont {Siu}, \bibfnamefont {K.~K.~W.}}, \bibinfo
  {author} {\bibnamefont {Hourigan}, \bibfnamefont {K.}}, \ and\ \bibinfo
  {author} {\bibnamefont {Fouras}, \bibfnamefont {A.}},\ }\bibfield  {title}
  {\enquote {\bibinfo {title} {Computed tomographic x-ray velocimetry for
  simultaneous 3d measurement of velocity and geometry in opaque vessels},}\
  }\href {\doibase 10.1007/s00348-010-1006-x} {\bibfield  {journal} {\bibinfo
  {journal} {Experiments in Fluids}\ }\textbf {\bibinfo {volume} {52}},\
  \bibinfo {pages} {543--554} (\bibinfo {year} {2012})}\BibitemShut {NoStop}%
\bibitem [{\citenamefont {Eyndhoven}\ \emph {et~al.}(2015)\citenamefont
  {Eyndhoven}, \citenamefont {Batenburg}, \citenamefont {Kazantsev},
  \citenamefont {Nieuwenhove}, \citenamefont {Lee}, \citenamefont {Dobson},\
  and\ \citenamefont {Sijbers}}]{VanEyndhoven2015}%
  \BibitemOpen
  \bibfield  {author} {\bibinfo {author} {\bibnamefont {Eyndhoven},
  \bibfnamefont {G.~V.}}, \bibinfo {author} {\bibnamefont {Batenburg},
  \bibfnamefont {K.~J.}}, \bibinfo {author} {\bibnamefont {Kazantsev},
  \bibfnamefont {D.}}, \bibinfo {author} {\bibnamefont {Nieuwenhove},
  \bibfnamefont {V.~V.}}, \bibinfo {author} {\bibnamefont {Lee}, \bibfnamefont
  {P.~D.}}, \bibinfo {author} {\bibnamefont {Dobson}, \bibfnamefont {K.~J.}}, \
  and\ \bibinfo {author} {\bibnamefont {Sijbers}, \bibfnamefont {J.}},\
  }\bibfield  {title} {\enquote {\bibinfo {title} {An iterative ct
  reconstruction algorithm for fast fluid flow imaging},}\ }\href {\doibase
  10.1109/TIP.2015.2466113} {\bibfield  {journal} {\bibinfo  {journal} {IEEE
  Transactions on Image Processing}\ }\textbf {\bibinfo {volume} {24}},\
  \bibinfo {pages} {4446--4458} (\bibinfo {year} {2015})}\BibitemShut {NoStop}%
\bibitem [{\citenamefont {Fouras}\ \emph {et~al.}(2007)\citenamefont {Fouras},
  \citenamefont {Dusting}, \citenamefont {Lewis},\ and\ \citenamefont
  {Hourigan}}]{Fouras2007}%
  \BibitemOpen
  \bibfield  {author} {\bibinfo {author} {\bibnamefont {Fouras}, \bibfnamefont
  {A.}}, \bibinfo {author} {\bibnamefont {Dusting}, \bibfnamefont {J.}},
  \bibinfo {author} {\bibnamefont {Lewis}, \bibfnamefont {R.}}, \ and\ \bibinfo
  {author} {\bibnamefont {Hourigan}, \bibfnamefont {K.}},\ }\bibfield  {title}
  {\enquote {\bibinfo {title} {Three-dimensional synchrotron x-ray particle
  image velocimetry},}\ }\href {\doibase 10.1063/1.2783978} {\bibfield
  {journal} {\bibinfo  {journal} {Journal of Applied Physics}\ }\textbf
  {\bibinfo {volume} {102}} (\bibinfo {year} {2007}),\
  10.1063/1.2783978}\BibitemShut {NoStop}%
\bibitem [{\citenamefont {Franchini}\ \emph {et~al.}(2019)\citenamefont
  {Franchini}, \citenamefont {Charogiannis}, \citenamefont {Markides},
  \citenamefont {Blunt},\ and\ \citenamefont {Krevor}}]{Franchini2019}%
  \BibitemOpen
  \bibfield  {author} {\bibinfo {author} {\bibnamefont {Franchini},
  \bibfnamefont {S.}}, \bibinfo {author} {\bibnamefont {Charogiannis},
  \bibfnamefont {A.}}, \bibinfo {author} {\bibnamefont {Markides},
  \bibfnamefont {C.~N.}}, \bibinfo {author} {\bibnamefont {Blunt},
  \bibfnamefont {M.~J.}}, \ and\ \bibinfo {author} {\bibnamefont {Krevor},
  \bibfnamefont {S.}},\ }\bibfield  {title} {\enquote {\bibinfo {title}
  {Advances in water resources calibration of astigmatic particle tracking
  velocimetry based on generalized gaussian feature extraction},}\ }\href
  {\doibase 10.1016/j.advwatres.2018.11.016} {\bibfield  {journal} {\bibinfo
  {journal} {Advances in Water Resources}\ }\textbf {\bibinfo {volume} {124}},\
  \bibinfo {pages} {1--8} (\bibinfo {year} {2019})}\BibitemShut {NoStop}%
\bibitem [{\citenamefont {Fu}, \citenamefont {Thomas},\ and\ \citenamefont
  {Li}(2021)}]{Fu2021}%
  \BibitemOpen
  \bibfield  {author} {\bibinfo {author} {\bibnamefont {Fu}, \bibfnamefont
  {J.}}, \bibinfo {author} {\bibnamefont {Thomas}, \bibfnamefont {H.~R.}}, \
  and\ \bibinfo {author} {\bibnamefont {Li}, \bibfnamefont {C.}},\ }\bibfield
  {title} {\enquote {\bibinfo {title} {Tortuosity of porous media: Image
  analysis and physical simulation},}\ }\href {\doibase
  10.1016/j.earscirev.2020.103439} {\bibfield  {journal} {\bibinfo  {journal}
  {Earth-Science Reviews}\ }\textbf {\bibinfo {volume} {212}},\ \bibinfo
  {pages} {103439} (\bibinfo {year} {2021})}\BibitemShut {NoStop}%
\bibitem [{\citenamefont {Gladden}\ and\ \citenamefont
  {Sederman}(2013)}]{Gladden2013}%
  \BibitemOpen
  \bibfield  {author} {\bibinfo {author} {\bibnamefont {Gladden}, \bibfnamefont
  {L.~F.}}\ and\ \bibinfo {author} {\bibnamefont {Sederman}, \bibfnamefont
  {A.~J.}},\ }\bibfield  {title} {\enquote {\bibinfo {title} {Recent advances
  in flow mri},}\ }\href {\doibase 10.1016/j.jmr.2012.11.022} {\bibfield
  {journal} {\bibinfo  {journal} {Journal of Magnetic Resonance}\ }\textbf
  {\bibinfo {volume} {229}},\ \bibinfo {pages} {2--11} (\bibinfo {year}
  {2013})}\BibitemShut {NoStop}%
\bibitem [{\citenamefont {Godinez}\ and\ \citenamefont
  {Rohr}(2015)}]{Godinez2015}%
  \BibitemOpen
  \bibfield  {author} {\bibinfo {author} {\bibnamefont {Godinez}, \bibfnamefont
  {W.~J.}}\ and\ \bibinfo {author} {\bibnamefont {Rohr}, \bibfnamefont {K.}},\
  }\bibfield  {title} {\enquote {\bibinfo {title} {Tracking multiple particles
  in fluorescence time-lapse microscopy images via probabilistic data
  association},}\ }\href {\doibase 10.1109/TMI.2014.2359541} {\bibfield
  {journal} {\bibinfo  {journal} {IEEE Transactions on Medical Imaging}\
  }\textbf {\bibinfo {volume} {34}},\ \bibinfo {pages} {415--432} (\bibinfo
  {year} {2015})}\BibitemShut {NoStop}%
\bibitem [{\citenamefont {Haffner}\ and\ \citenamefont
  {Mirbod}(2020)}]{Haffner2020}%
  \BibitemOpen
  \bibfield  {author} {\bibinfo {author} {\bibnamefont {Haffner}, \bibfnamefont
  {E.~A.}}\ and\ \bibinfo {author} {\bibnamefont {Mirbod}, \bibfnamefont
  {P.}},\ }\bibfield  {title} {\enquote {\bibinfo {title} {Velocity
  measurements of dilute particulate suspension over and through a porous
  medium model},}\ }\href {\doibase 10.1063/5.0015207} {\bibfield  {journal}
  {\bibinfo  {journal} {Physics of Fluids}\ }\textbf {\bibinfo {volume} {32}},\
  \bibinfo {pages} {083608} (\bibinfo {year} {2020})}\BibitemShut {NoStop}%
\bibitem [{\citenamefont {Heyndrickx}\ \emph {et~al.}(2020)\citenamefont
  {Heyndrickx}, \citenamefont {Bultreys}, \citenamefont {Goethals},
  \citenamefont {Hoorebeke},\ and\ \citenamefont {Boone}}]{Heyndrickx2020}%
  \BibitemOpen
  \bibfield  {author} {\bibinfo {author} {\bibnamefont {Heyndrickx},
  \bibfnamefont {M.}}, \bibinfo {author} {\bibnamefont {Bultreys},
  \bibfnamefont {T.}}, \bibinfo {author} {\bibnamefont {Goethals},
  \bibfnamefont {W.}}, \bibinfo {author} {\bibnamefont {Hoorebeke},
  \bibfnamefont {L.~V.}}, \ and\ \bibinfo {author} {\bibnamefont {Boone},
  \bibfnamefont {M.~N.}},\ }\bibfield  {title} {\enquote {\bibinfo {title}
  {Improving image quality in fast, time-resolved micro-ct by weighted back
  projection},}\ }\href {\doibase 10.1038/s41598-020-74827-x} {\bibfield
  {journal} {\bibinfo  {journal} {Scientific Reports}\ }\textbf {\bibinfo
  {volume} {10}},\ \bibinfo {pages} {18029} (\bibinfo {year}
  {2020})}\BibitemShut {NoStop}%
\bibitem [{\citenamefont {Holtzman}(2016)}]{Holtzman2016}%
  \BibitemOpen
  \bibfield  {author} {\bibinfo {author} {\bibnamefont {Holtzman},
  \bibfnamefont {R.}},\ }\bibfield  {title} {\enquote {\bibinfo {title}
  {Effects of pore-scale disorder on fluid displacement in partially-wettable
  porous media},}\ }\href {\doibase 10.1038/srep36221} {\bibfield  {journal}
  {\bibinfo  {journal} {Scientific Reports}\ }\textbf {\bibinfo {volume} {6}},\
  \bibinfo {pages} {36221} (\bibinfo {year} {2016})}\BibitemShut {NoStop}%
\bibitem [{\citenamefont {Jaqaman}\ \emph {et~al.}(2008)\citenamefont
  {Jaqaman}, \citenamefont {Loerke}, \citenamefont {Mettlen}, \citenamefont
  {Kuwata}, \citenamefont {Grinstein}, \citenamefont {Schmid},\ and\
  \citenamefont {Danuser}}]{Jaqaman2008}%
  \BibitemOpen
  \bibfield  {author} {\bibinfo {author} {\bibnamefont {Jaqaman}, \bibfnamefont
  {K.}}, \bibinfo {author} {\bibnamefont {Loerke}, \bibfnamefont {D.}},
  \bibinfo {author} {\bibnamefont {Mettlen}, \bibfnamefont {M.}}, \bibinfo
  {author} {\bibnamefont {Kuwata}, \bibfnamefont {H.}}, \bibinfo {author}
  {\bibnamefont {Grinstein}, \bibfnamefont {S.}}, \bibinfo {author}
  {\bibnamefont {Schmid}, \bibfnamefont {S.~L.}}, \ and\ \bibinfo {author}
  {\bibnamefont {Danuser}, \bibfnamefont {G.}},\ }\bibfield  {title} {\enquote
  {\bibinfo {title} {Robust single-particle tracking in live-cell time-lapse
  sequences},}\ }\href {\doibase 10.1038/nmeth.1237} {\bibfield  {journal}
  {\bibinfo  {journal} {Nature Methods}\ }\textbf {\bibinfo {volume} {5}},\
  \bibinfo {pages} {695--702} (\bibinfo {year} {2008})}\BibitemShut {NoStop}%
\bibitem [{\citenamefont {de~Kort}\ \emph {et~al.}(2019)\citenamefont
  {de~Kort}, \citenamefont {Hertel}, \citenamefont {Appel}, \citenamefont
  {de~Jong}, \citenamefont {Mantle}, \citenamefont {Sederman},\ and\
  \citenamefont {Gladden}}]{DeKort2019}%
  \BibitemOpen
  \bibfield  {author} {\bibinfo {author} {\bibnamefont {de~Kort}, \bibfnamefont
  {D.~W.}}, \bibinfo {author} {\bibnamefont {Hertel}, \bibfnamefont {S.~A.}},
  \bibinfo {author} {\bibnamefont {Appel}, \bibfnamefont {M.}}, \bibinfo
  {author} {\bibnamefont {de~Jong}, \bibfnamefont {H.}}, \bibinfo {author}
  {\bibnamefont {Mantle}, \bibfnamefont {M.~D.}}, \bibinfo {author}
  {\bibnamefont {Sederman}, \bibfnamefont {A.~J.}}, \ and\ \bibinfo {author}
  {\bibnamefont {Gladden}, \bibfnamefont {L.~F.}},\ }\bibfield  {title}
  {\enquote {\bibinfo {title} {Under-sampling and compressed sensing of 3d
  spatially-resolved displacement propagators in porous media using apgste-rare
  mri},}\ }\href {\doibase 10.1016/j.mri.2018.08.014} {\bibfield  {journal}
  {\bibinfo  {journal} {Magnetic Resonance Imaging}\ }\textbf {\bibinfo
  {volume} {56}},\ \bibinfo {pages} {24--31} (\bibinfo {year}
  {2019})}\BibitemShut {NoStop}%
\bibitem [{\citenamefont {Lee}\ and\ \citenamefont {Kim}(2003)}]{Lee2003}%
  \BibitemOpen
  \bibfield  {author} {\bibinfo {author} {\bibnamefont {Lee}, \bibfnamefont
  {S.-J.}}\ and\ \bibinfo {author} {\bibnamefont {Kim}, \bibfnamefont
  {G.-B.}},\ }\bibfield  {title} {\enquote {\bibinfo {title} {X-ray particle
  image velocimetry for measuring quantitative flow information inside opaque
  objects},}\ }\href {\doibase 10.1063/1.1599981} {\bibfield  {journal}
  {\bibinfo  {journal} {Journal of Applied Physics}\ }\textbf {\bibinfo
  {volume} {94}},\ \bibinfo {pages} {3620--3623} (\bibinfo {year}
  {2003})}\BibitemShut {NoStop}%
\bibitem [{\citenamefont {Lenormand}, \citenamefont {Zarcone},\ and\
  \citenamefont {Sarr}(1983)}]{Lenormand1983}%
  \BibitemOpen
  \bibfield  {author} {\bibinfo {author} {\bibnamefont {Lenormand},
  \bibfnamefont {R.}}, \bibinfo {author} {\bibnamefont {Zarcone}, \bibfnamefont
  {C.}}, \ and\ \bibinfo {author} {\bibnamefont {Sarr}, \bibfnamefont {A.}},\
  }\bibfield  {title} {\enquote {\bibinfo {title} {Mechanisms of the
  displacement of one fluid by another in a network of capillary ducts},}\
  }\href {\doibase 10.1017/S0022112083003110} {\bibfield  {journal} {\bibinfo
  {journal} {Journal of Fluid Mechanics}\ }\textbf {\bibinfo {volume} {135}},\
  \bibinfo {pages} {337} (\bibinfo {year} {1983})}\BibitemShut {NoStop}%
\bibitem [{\citenamefont {Ling}\ \emph {et~al.}(2017)\citenamefont {Ling},
  \citenamefont {Bao}, \citenamefont {Oostrom}, \citenamefont {Battiato},\ and\
  \citenamefont {Tartakovsky}}]{Ling2017}%
  \BibitemOpen
  \bibfield  {author} {\bibinfo {author} {\bibnamefont {Ling}, \bibfnamefont
  {B.}}, \bibinfo {author} {\bibnamefont {Bao}, \bibfnamefont {J.}}, \bibinfo
  {author} {\bibnamefont {Oostrom}, \bibfnamefont {M.}}, \bibinfo {author}
  {\bibnamefont {Battiato}, \bibfnamefont {I.}}, \ and\ \bibinfo {author}
  {\bibnamefont {Tartakovsky}, \bibfnamefont {A.~M.}},\ }\bibfield  {title}
  {\enquote {\bibinfo {title} {Modeling variability in porescale multiphase
  flow experiments},}\ }\href {\doibase 10.1016/j.advwatres.2017.04.005}
  {\bibfield  {journal} {\bibinfo  {journal} {Advances in Water Resources}\
  }\textbf {\bibinfo {volume} {105}},\ \bibinfo {pages} {29--38} (\bibinfo
  {year} {2017})}\BibitemShut {NoStop}%
\bibitem [{\citenamefont {Mascini}\ \emph {et~al.}(2021)\citenamefont
  {Mascini}, \citenamefont {Boone}, \citenamefont {Offenwert}, \citenamefont
  {Wang}, \citenamefont {Cnudde},\ and\ \citenamefont
  {Bultreys}}]{Mascini2021a}%
  \BibitemOpen
  \bibfield  {author} {\bibinfo {author} {\bibnamefont {Mascini}, \bibfnamefont
  {A.}}, \bibinfo {author} {\bibnamefont {Boone}, \bibfnamefont {M.}}, \bibinfo
  {author} {\bibnamefont {Offenwert}, \bibfnamefont {S.~V.}}, \bibinfo {author}
  {\bibnamefont {Wang}, \bibfnamefont {S.}}, \bibinfo {author} {\bibnamefont
  {Cnudde}, \bibfnamefont {V.}}, \ and\ \bibinfo {author} {\bibnamefont
  {Bultreys}, \bibfnamefont {T.}},\ }\bibfield  {title} {\enquote {\bibinfo
  {title} {Fluid invasion dynamics in porous media with complex wettability and
  connectivity},}\ }\href {\doibase 10.1029/2021GL095185} {\bibfield  {journal}
  {\bibinfo  {journal} {Geophysical Research Letters}\ }\textbf {\bibinfo
  {volume} {48}} (\bibinfo {year} {2021}),\ 10.1029/2021GL095185}\BibitemShut
  {NoStop}%
\bibitem [{\citenamefont {McClure}, \citenamefont {Berg},\ and\ \citenamefont
  {Armstrong}(2021)}]{McClure2021}%
  \BibitemOpen
  \bibfield  {author} {\bibinfo {author} {\bibnamefont {McClure}, \bibfnamefont
  {J.~E.}}, \bibinfo {author} {\bibnamefont {Berg}, \bibfnamefont {S.}}, \ and\
  \bibinfo {author} {\bibnamefont {Armstrong}, \bibfnamefont {R.~T.}},\
  }\bibfield  {title} {\enquote {\bibinfo {title} {Capillary fluctuations and
  energy dynamics for flow in porous media},}\ }\href {\doibase
  10.1063/5.0057428} {\bibfield  {journal} {\bibinfo  {journal} {Physics of
  Fluids}\ }\textbf {\bibinfo {volume} {083323}},\ \bibinfo {pages} {1--16}
  (\bibinfo {year} {2021})}\BibitemShut {NoStop}%
\bibitem [{\citenamefont {Melling}(1997)}]{Melling1997}%
  \BibitemOpen
  \bibfield  {author} {\bibinfo {author} {\bibnamefont {Melling}, \bibfnamefont
  {A.}},\ }\bibfield  {title} {\enquote {\bibinfo {title} {Tracer particles and
  seeding for particle image velocimetry},}\ }\href {\doibase
  10.1088/0957-0233/8/12/005} {\bibfield  {journal} {\bibinfo  {journal}
  {Measurement Science and Technology}\ }\textbf {\bibinfo {volume} {8}},\
  \bibinfo {pages} {1406--1416} (\bibinfo {year} {1997})}\BibitemShut {NoStop}%
\bibitem [{\citenamefont {Mercer}\ and\ \citenamefont
  {Cohen}(1990)}]{Mercer1990}%
  \BibitemOpen
  \bibfield  {author} {\bibinfo {author} {\bibnamefont {Mercer}, \bibfnamefont
  {J.~W.}}\ and\ \bibinfo {author} {\bibnamefont {Cohen}, \bibfnamefont
  {R.~M.}},\ }\bibfield  {title} {\enquote {\bibinfo {title} {A review of
  immiscible fluids in the subsurface: Properties, models, characterization and
  remediation},}\ }\href {\doibase 10.1016/0169-7722(90)90043-G} {\bibfield
  {journal} {\bibinfo  {journal} {Journal of Contaminant Hydrology}\ }\textbf
  {\bibinfo {volume} {6}},\ \bibinfo {pages} {107--163} (\bibinfo {year}
  {1990})}\BibitemShut {NoStop}%
\bibitem [{\citenamefont {Miele}, \citenamefont {Anna},\ and\ \citenamefont
  {Dentz}(2019)}]{Miele2019}%
  \BibitemOpen
  \bibfield  {author} {\bibinfo {author} {\bibnamefont {Miele}, \bibfnamefont
  {F.}}, \bibinfo {author} {\bibnamefont {Anna}, \bibfnamefont {P.~D.}}, \ and\
  \bibinfo {author} {\bibnamefont {Dentz}, \bibfnamefont {M.}},\ }\bibfield
  {title} {\enquote {\bibinfo {title} {Stochastic model for filtration by
  porous materials},}\ }\href {\doibase 10.1103/PhysRevFluids.4.094101}
  {\bibfield  {journal} {\bibinfo  {journal} {Physical Review Fluids}\ }\textbf
  {\bibinfo {volume} {4}},\ \bibinfo {pages} {94101} (\bibinfo {year}
  {2019})}\BibitemShut {NoStop}%
\bibitem [{\citenamefont {Molnar}\ \emph {et~al.}(2015)\citenamefont {Molnar},
  \citenamefont {Johnson}, \citenamefont {Gerhard}, \citenamefont {Willson},\
  and\ \citenamefont {O'Carroll}}]{Molnar2015}%
  \BibitemOpen
  \bibfield  {author} {\bibinfo {author} {\bibnamefont {Molnar}, \bibfnamefont
  {I.~L.}}, \bibinfo {author} {\bibnamefont {Johnson}, \bibfnamefont {W.~P.}},
  \bibinfo {author} {\bibnamefont {Gerhard}, \bibfnamefont {J.~I.}}, \bibinfo
  {author} {\bibnamefont {Willson}, \bibfnamefont {C.~S.}}, \ and\ \bibinfo
  {author} {\bibnamefont {O'Carroll}, \bibfnamefont {D.~M.}},\ }\bibfield
  {title} {\enquote {\bibinfo {title} {Predicting colloid transport through
  saturated porous media: A critical review},}\ }\href {\doibase
  10.1002/2015WR017318} {\bibfield  {journal} {\bibinfo  {journal} {Water
  Resources Research}\ }\textbf {\bibinfo {volume} {51}},\ \bibinfo {pages}
  {6804--6845} (\bibinfo {year} {2015})}\BibitemShut {NoStop}%
\bibitem [{\citenamefont {Mouli-Castillo}\ \emph {et~al.}(2019)\citenamefont
  {Mouli-Castillo}, \citenamefont {Wilkinson}, \citenamefont {Mignard},
  \citenamefont {McDermott}, \citenamefont {Haszeldine},\ and\ \citenamefont
  {Shipton}}]{Mouli-Castillo2019}%
  \BibitemOpen
  \bibfield  {author} {\bibinfo {author} {\bibnamefont {Mouli-Castillo},
  \bibfnamefont {J.}}, \bibinfo {author} {\bibnamefont {Wilkinson},
  \bibfnamefont {M.}}, \bibinfo {author} {\bibnamefont {Mignard}, \bibfnamefont
  {D.}}, \bibinfo {author} {\bibnamefont {McDermott}, \bibfnamefont {C.}},
  \bibinfo {author} {\bibnamefont {Haszeldine}, \bibfnamefont {R.~S.}}, \ and\
  \bibinfo {author} {\bibnamefont {Shipton}, \bibfnamefont {Z.~K.}},\
  }\bibfield  {title} {\enquote {\bibinfo {title} {Inter-seasonal
  compressed-air energy storage using saline aquifers},}\ }\href {\doibase
  10.1038/s41560-018-0311-0} {\bibfield  {journal} {\bibinfo  {journal} {Nature
  Energy}\ }\textbf {\bibinfo {volume} {4}},\ \bibinfo {pages} {131--139}
  (\bibinfo {year} {2019})}\BibitemShut {NoStop}%
\bibitem [{\citenamefont {Mularczyk}\ \emph {et~al.}(2020)\citenamefont
  {Mularczyk}, \citenamefont {Lin}, \citenamefont {Blunt}, \citenamefont
  {Lamibrac}, \citenamefont {Marone}, \citenamefont {Schmidt}, \citenamefont
  {Büchi},\ and\ \citenamefont {Eller}}]{Mularczyk2020}%
  \BibitemOpen
  \bibfield  {author} {\bibinfo {author} {\bibnamefont {Mularczyk},
  \bibfnamefont {A.}}, \bibinfo {author} {\bibnamefont {Lin}, \bibfnamefont
  {Q.}}, \bibinfo {author} {\bibnamefont {Blunt}, \bibfnamefont {M.~J.}},
  \bibinfo {author} {\bibnamefont {Lamibrac}, \bibfnamefont {A.}}, \bibinfo
  {author} {\bibnamefont {Marone}, \bibfnamefont {F.}}, \bibinfo {author}
  {\bibnamefont {Schmidt}, \bibfnamefont {T.~J.}}, \bibinfo {author}
  {\bibnamefont {Büchi}, \bibfnamefont {F.~N.}}, \ and\ \bibinfo {author}
  {\bibnamefont {Eller}, \bibfnamefont {J.}},\ }\bibfield  {title} {\enquote
  {\bibinfo {title} {Droplet and percolation network interactions in a fuel
  cell gas diffusion layer},}\ }\href {\doibase 10.1149/1945-7111/ab8c85}
  {\bibfield  {journal} {\bibinfo  {journal} {Journal of The Electrochemical
  Society}\ }\textbf {\bibinfo {volume} {167}},\ \bibinfo {pages} {084506}
  (\bibinfo {year} {2020})}\BibitemShut {NoStop}%
\bibitem [{\citenamefont {Myers}\ \emph {et~al.}(2011)\citenamefont {Myers},
  \citenamefont {Kingston}, \citenamefont {Varslot}, \citenamefont {Turner},\
  and\ \citenamefont {Sheppard}}]{Myers2011}%
  \BibitemOpen
  \bibfield  {author} {\bibinfo {author} {\bibnamefont {Myers}, \bibfnamefont
  {G.~R.}}, \bibinfo {author} {\bibnamefont {Kingston}, \bibfnamefont {A.~M.}},
  \bibinfo {author} {\bibnamefont {Varslot}, \bibfnamefont {T.~K.}}, \bibinfo
  {author} {\bibnamefont {Turner}, \bibfnamefont {M.~L.}}, \ and\ \bibinfo
  {author} {\bibnamefont {Sheppard}, \bibfnamefont {A.~P.}},\ }\bibfield
  {title} {\enquote {\bibinfo {title} {Dynamic x-ray micro-tomography for real
  time imaging of drainage and imbibition processes at the pore scale},}\ \
  }(\bibinfo  {publisher} {Society of Core Analysts},\ \bibinfo {year}
  {2011})\BibitemShut {NoStop}%
\bibitem [{\citenamefont {Mäkiharju}\ \emph {et~al.}(2022)\citenamefont
  {Mäkiharju}, \citenamefont {Dewanckele}, \citenamefont {Boone},
  \citenamefont {Wagner},\ and\ \citenamefont {Griesser}}]{Makiharju2022}%
  \BibitemOpen
  \bibfield  {author} {\bibinfo {author} {\bibnamefont {Mäkiharju},
  \bibfnamefont {S.~A.}}, \bibinfo {author} {\bibnamefont {Dewanckele},
  \bibfnamefont {J.}}, \bibinfo {author} {\bibnamefont {Boone}, \bibfnamefont
  {M.}}, \bibinfo {author} {\bibnamefont {Wagner}, \bibfnamefont {C.}}, \ and\
  \bibinfo {author} {\bibnamefont {Griesser}, \bibfnamefont {A.}},\ }\bibfield
  {title} {\enquote {\bibinfo {title} {Tomographic x-ray particle tracking
  velocimetry},}\ }\href {\doibase 10.1007/s00348-021-03362-w} {\bibfield
  {journal} {\bibinfo  {journal} {Experiments in Fluids}\ }\textbf {\bibinfo
  {volume} {63}},\ \bibinfo {pages} {16} (\bibinfo {year} {2022})}\BibitemShut
  {NoStop}%
\bibitem [{\citenamefont {Ouellette}, \citenamefont {Xu},\ and\ \citenamefont
  {Bodenschatz}(2006)}]{Ouellette2006}%
  \BibitemOpen
  \bibfield  {author} {\bibinfo {author} {\bibnamefont {Ouellette},
  \bibfnamefont {N.~T.}}, \bibinfo {author} {\bibnamefont {Xu}, \bibfnamefont
  {H.}}, \ and\ \bibinfo {author} {\bibnamefont {Bodenschatz}, \bibfnamefont
  {E.}},\ }\bibfield  {title} {\enquote {\bibinfo {title} {A quantitative study
  of three-dimensional lagrangian particle tracking algorithms},}\ }\href
  {\doibase 10.1007/s00348-005-0068-7} {\bibfield  {journal} {\bibinfo
  {journal} {Experiments in Fluids}\ }\textbf {\bibinfo {volume} {40}},\
  \bibinfo {pages} {301--313} (\bibinfo {year} {2006})}\BibitemShut {NoStop}%
\bibitem [{\citenamefont {Primkulov}\ \emph {et~al.}(2019)\citenamefont
  {Primkulov}, \citenamefont {Pahlavan}, \citenamefont {Fu}, \citenamefont
  {Zhao}, \citenamefont {MacMinn},\ and\ \citenamefont
  {Juanes}}]{Primkulov2019}%
  \BibitemOpen
  \bibfield  {author} {\bibinfo {author} {\bibnamefont {Primkulov},
  \bibfnamefont {B.~K.}}, \bibinfo {author} {\bibnamefont {Pahlavan},
  \bibfnamefont {A.~A.}}, \bibinfo {author} {\bibnamefont {Fu}, \bibfnamefont
  {X.}}, \bibinfo {author} {\bibnamefont {Zhao}, \bibfnamefont {B.}}, \bibinfo
  {author} {\bibnamefont {MacMinn}, \bibfnamefont {C.~W.}}, \ and\ \bibinfo
  {author} {\bibnamefont {Juanes}, \bibfnamefont {R.}},\ }\bibfield  {title}
  {\enquote {\bibinfo {title} {Signatures of fluid–fluid displacement in
  porous media: wettability, patterns and pressures},}\ }\href {\doibase
  10.1017/jfm.2019.554} {\bibfield  {journal} {\bibinfo  {journal} {Journal of
  Fluid Mechanics}\ }\textbf {\bibinfo {volume} {875}},\ \bibinfo {pages} {R4}
  (\bibinfo {year} {2019})}\BibitemShut {NoStop}%
\bibitem [{\citenamefont {Raeini}\ and\ \citenamefont
  {Blunt}(2022)}]{Raeini2022}%
  \BibitemOpen
  \bibfield  {author} {\bibinfo {author} {\bibnamefont {Raeini}, \bibfnamefont
  {A.}}\ and\ \bibinfo {author} {\bibnamefont {Blunt}, \bibfnamefont {M.}},\
  }\href@noop {} {\enquote {\bibinfo {title} {Imperial college london
  pore-scale modelling and imaging github page},}\ } (\bibinfo {year}
  {2022})\BibitemShut {NoStop}%
\bibitem [{\citenamefont {Raeini}, \citenamefont {Bijeljic},\ and\
  \citenamefont {Blunt}(2017)}]{Raeini2017}%
  \BibitemOpen
  \bibfield  {author} {\bibinfo {author} {\bibnamefont {Raeini}, \bibfnamefont
  {A.~Q.}}, \bibinfo {author} {\bibnamefont {Bijeljic}, \bibfnamefont {B.}}, \
  and\ \bibinfo {author} {\bibnamefont {Blunt}, \bibfnamefont {M.~J.}},\
  }\bibfield  {title} {\enquote {\bibinfo {title} {Generalized network
  modeling: Network extraction as a coarse-scale discretization of the void
  space of porous media},}\ }\href {\doibase 10.1103/PhysRevE.96.013312}
  {\bibfield  {journal} {\bibinfo  {journal} {Physical Review E}\ }\textbf
  {\bibinfo {volume} {96}},\ \bibinfo {pages} {013312} (\bibinfo {year}
  {2017})}\BibitemShut {NoStop}%
\bibitem [{\citenamefont {Raffael}\ \emph {et~al.}(2018)\citenamefont
  {Raffael}, \citenamefont {Willert}, \citenamefont {Wereley},\ and\
  \citenamefont {Kompenhans}}]{Raffael2018}%
  \BibitemOpen
  \bibfield  {author} {\bibinfo {author} {\bibnamefont {Raffael}, \bibfnamefont
  {M.}}, \bibinfo {author} {\bibnamefont {Willert}, \bibfnamefont {C.}},
  \bibinfo {author} {\bibnamefont {Wereley}, \bibfnamefont {S.~T.}}, \ and\
  \bibinfo {author} {\bibnamefont {Kompenhans}, \bibfnamefont {J.}},\ }\href
  {\doibase 10.1007/978-3-540-72308-0} {\emph {\bibinfo {title} {Particle Image
  Velocimetry (the Third Edition)}}}\ (\bibinfo {year} {2018})\ p.\ \bibinfo
  {pages} {680}\BibitemShut {NoStop}%
\bibitem [{\citenamefont {Roman}\ \emph {et~al.}(2015)\citenamefont {Roman},
  \citenamefont {Soulaine}, \citenamefont {AlSaud}, \citenamefont {Kovscek},\
  and\ \citenamefont {Tchelepi}}]{Roman2015}%
  \BibitemOpen
  \bibfield  {author} {\bibinfo {author} {\bibnamefont {Roman}, \bibfnamefont
  {S.}}, \bibinfo {author} {\bibnamefont {Soulaine}, \bibfnamefont {C.}},
  \bibinfo {author} {\bibnamefont {AlSaud}, \bibfnamefont {M.~A.}}, \bibinfo
  {author} {\bibnamefont {Kovscek}, \bibfnamefont {A.}}, \ and\ \bibinfo
  {author} {\bibnamefont {Tchelepi}, \bibfnamefont {H.}},\ }\bibfield  {title}
  {\enquote {\bibinfo {title} {Particle velocimetry analysis of immiscible
  two-phase flow in micromodels},}\ }\href {\doibase
  10.1016/j.advwatres.2015.08.015} {\bibfield  {journal} {\bibinfo  {journal}
  {Advances in Water Resources}\ }\textbf {\bibinfo {volume} {000}},\ \bibinfo
  {pages} {1--13} (\bibinfo {year} {2015})}\BibitemShut {NoStop}%
\bibitem [{\citenamefont {Russell}\ and\ \citenamefont
  {Bedrikovetsky}(2021)}]{Russell2021}%
  \BibitemOpen
  \bibfield  {author} {\bibinfo {author} {\bibnamefont {Russell}, \bibfnamefont
  {T.}}\ and\ \bibinfo {author} {\bibnamefont {Bedrikovetsky}, \bibfnamefont
  {P.}},\ }\bibfield  {title} {\enquote {\bibinfo {title} {Boltzmann's
  colloidal transport in porous media with velocity-dependent capture
  probability},}\ }\href {\doibase 10.1063/5.0035392} {\bibfield  {journal}
  {\bibinfo  {journal} {Physics of Fluids}\ }\textbf {\bibinfo {volume} {33}},\
  \bibinfo {pages} {053306} (\bibinfo {year} {2021})}\BibitemShut {NoStop}%
\bibitem [{\citenamefont {Saxena}\ \emph {et~al.}(2017)\citenamefont {Saxena},
  \citenamefont {Hofmann}, \citenamefont {Alpak}, \citenamefont {Berg},
  \citenamefont {Dietderich}, \citenamefont {Agarwal}, \citenamefont {Tandon},
  \citenamefont {Hunter}, \citenamefont {Freeman},\ and\ \citenamefont
  {Wilson}}]{Saxena2017}%
  \BibitemOpen
  \bibfield  {author} {\bibinfo {author} {\bibnamefont {Saxena}, \bibfnamefont
  {N.}}, \bibinfo {author} {\bibnamefont {Hofmann}, \bibfnamefont {R.}},
  \bibinfo {author} {\bibnamefont {Alpak}, \bibfnamefont {F.~O.}}, \bibinfo
  {author} {\bibnamefont {Berg}, \bibfnamefont {S.}}, \bibinfo {author}
  {\bibnamefont {Dietderich}, \bibfnamefont {J.}}, \bibinfo {author}
  {\bibnamefont {Agarwal}, \bibfnamefont {U.}}, \bibinfo {author} {\bibnamefont
  {Tandon}, \bibfnamefont {K.}}, \bibinfo {author} {\bibnamefont {Hunter},
  \bibfnamefont {S.}}, \bibinfo {author} {\bibnamefont {Freeman}, \bibfnamefont
  {J.}}, \ and\ \bibinfo {author} {\bibnamefont {Wilson}, \bibfnamefont
  {O.~B.}},\ }\bibfield  {title} {\enquote {\bibinfo {title} {References and
  benchmarks for pore-scale flow simulated using micro-ct images of porous
  media and digital rocks},}\ }\href {\doibase 10.1016/j.advwatres.2017.09.007}
  {\bibfield  {journal} {\bibinfo  {journal} {Advances in Water Resources}\
  }\textbf {\bibinfo {volume} {109}},\ \bibinfo {pages} {211--235} (\bibinfo
  {year} {2017})}\BibitemShut {NoStop}%
\bibitem [{\citenamefont {Schanz}, \citenamefont {Gesemann},\ and\
  \citenamefont {Schröder}(2016)}]{Schanz2016}%
  \BibitemOpen
  \bibfield  {author} {\bibinfo {author} {\bibnamefont {Schanz}, \bibfnamefont
  {D.}}, \bibinfo {author} {\bibnamefont {Gesemann}, \bibfnamefont {S.}}, \
  and\ \bibinfo {author} {\bibnamefont {Schröder}, \bibfnamefont {A.}},\
  }\bibfield  {title} {\enquote {\bibinfo {title} {Shake-the-box: Lagrangian
  particle tracking at high particle image densities},}\ }\href {\doibase
  10.1007/s00348-016-2157-1} {\bibfield  {journal} {\bibinfo  {journal}
  {Experiments in Fluids}\ }\textbf {\bibinfo {volume} {57}},\ \bibinfo {pages}
  {1--27} (\bibinfo {year} {2016})}\BibitemShut {NoStop}%
\bibitem [{\citenamefont {Schryver}\ \emph {et~al.}(2018)\citenamefont
  {Schryver}, \citenamefont {Dierick}, \citenamefont {Heyndrickx},
  \citenamefont {Stappen}, \citenamefont {Boone}, \citenamefont {Hoorebeke},\
  and\ \citenamefont {Boone}}]{DeSchryver2018}%
  \BibitemOpen
  \bibfield  {author} {\bibinfo {author} {\bibnamefont {Schryver},
  \bibfnamefont {T.~D.}}, \bibinfo {author} {\bibnamefont {Dierick},
  \bibfnamefont {M.}}, \bibinfo {author} {\bibnamefont {Heyndrickx},
  \bibfnamefont {M.}}, \bibinfo {author} {\bibnamefont {Stappen}, \bibfnamefont
  {J.~V.}}, \bibinfo {author} {\bibnamefont {Boone}, \bibfnamefont {M.~A.}},
  \bibinfo {author} {\bibnamefont {Hoorebeke}, \bibfnamefont {L.~V.}}, \ and\
  \bibinfo {author} {\bibnamefont {Boone}, \bibfnamefont {M.~N.}},\ }\bibfield
  {title} {\enquote {\bibinfo {title} {Motion compensated micro-ct
  reconstruction for in-situ analysis of dynamic processes},}\ }\href {\doibase
  10.1038/s41598-018-25916-5} {\bibfield  {journal} {\bibinfo  {journal}
  {Scientific Reports}\ }\textbf {\bibinfo {volume} {8}},\ \bibinfo {pages}
  {7655} (\bibinfo {year} {2018})}\BibitemShut {NoStop}%
\bibitem [{\citenamefont {Segur}\ and\ \citenamefont
  {Oberstar}(1951)}]{Segur1951}%
  \BibitemOpen
  \bibfield  {author} {\bibinfo {author} {\bibnamefont {Segur}, \bibfnamefont
  {J.~B.}}\ and\ \bibinfo {author} {\bibnamefont {Oberstar}, \bibfnamefont
  {H.~E.}},\ }\bibfield  {title} {\enquote {\bibinfo {title} {Viscosity of
  glycerol and its aqueous solutions},}\ }\href {\doibase 10.1021/ie50501a040}
  {\bibfield  {journal} {\bibinfo  {journal} {Industrial and Engineering
  Chemistry}\ }\textbf {\bibinfo {volume} {43}},\ \bibinfo {pages} {2117--2120}
  (\bibinfo {year} {1951})}\BibitemShut {NoStop}%
\bibitem [{\citenamefont {Singh}\ \emph {et~al.}(2019)\citenamefont {Singh},
  \citenamefont {Jung}, \citenamefont {Brinkmann},\ and\ \citenamefont
  {Seemann}}]{Singh2019b}%
  \BibitemOpen
  \bibfield  {author} {\bibinfo {author} {\bibnamefont {Singh}, \bibfnamefont
  {K.}}, \bibinfo {author} {\bibnamefont {Jung}, \bibfnamefont {M.}}, \bibinfo
  {author} {\bibnamefont {Brinkmann}, \bibfnamefont {M.}}, \ and\ \bibinfo
  {author} {\bibnamefont {Seemann}, \bibfnamefont {R.}},\ }\bibfield  {title}
  {\enquote {\bibinfo {title} {Capillary-dominated fluid displacement in porous
  media},}\ }\href {\doibase 10.1146/annurev-fluid-010518-040342} {\bibfield
  {journal} {\bibinfo  {journal} {Annual Review of Fluid Mechanics}\ }\textbf
  {\bibinfo {volume} {51}},\ \bibinfo {pages} {429--449} (\bibinfo {year}
  {2019})}\BibitemShut {NoStop}%
\bibitem [{\citenamefont {Spurin}\ \emph {et~al.}(2021)\citenamefont {Spurin},
  \citenamefont {Bultreys}, \citenamefont {Rücker}, \citenamefont {Garfi},
  \citenamefont {Schlepütz}, \citenamefont {Novak}, \citenamefont {Berg},
  \citenamefont {Blunt},\ and\ \citenamefont {Krevor}}]{Spurin2021}%
  \BibitemOpen
  \bibfield  {author} {\bibinfo {author} {\bibnamefont {Spurin}, \bibfnamefont
  {C.}}, \bibinfo {author} {\bibnamefont {Bultreys}, \bibfnamefont {T.}},
  \bibinfo {author} {\bibnamefont {Rücker}, \bibfnamefont {M.}}, \bibinfo
  {author} {\bibnamefont {Garfi}, \bibfnamefont {G.}}, \bibinfo {author}
  {\bibnamefont {Schlepütz}, \bibfnamefont {C.~M.}}, \bibinfo {author}
  {\bibnamefont {Novak}, \bibfnamefont {V.}}, \bibinfo {author} {\bibnamefont
  {Berg}, \bibfnamefont {S.}}, \bibinfo {author} {\bibnamefont {Blunt},
  \bibfnamefont {M.~J.}}, \ and\ \bibinfo {author} {\bibnamefont {Krevor},
  \bibfnamefont {S.}},\ }\bibfield  {title} {\enquote {\bibinfo {title} {The
  development of intermittent multiphase fluid flow pathways through a porous
  rock},}\ }\href {\doibase 10.1016/j.advwatres.2021.103868} {\bibfield
  {journal} {\bibinfo  {journal} {Advances in Water Resources}\ }\textbf
  {\bibinfo {volume} {150}},\ \bibinfo {pages} {103868} (\bibinfo {year}
  {2021})}\BibitemShut {NoStop}%
\bibitem [{\citenamefont {Takamura}, \citenamefont {Fischer},\ and\
  \citenamefont {Morrow}(2012)}]{Takamura2012}%
  \BibitemOpen
  \bibfield  {author} {\bibinfo {author} {\bibnamefont {Takamura},
  \bibfnamefont {K.}}, \bibinfo {author} {\bibnamefont {Fischer}, \bibfnamefont
  {H.}}, \ and\ \bibinfo {author} {\bibnamefont {Morrow}, \bibfnamefont
  {N.~R.}},\ }\bibfield  {title} {\enquote {\bibinfo {title} {Physical
  properties of aqueous glycerol solutions},}\ }\href {\doibase
  10.1016/j.petrol.2012.09.003} {\bibfield  {journal} {\bibinfo  {journal}
  {Journal of Petroleum Science and Engineering}\ }\textbf {\bibinfo {volume}
  {98-99}},\ \bibinfo {pages} {50--60} (\bibinfo {year} {2012})}\BibitemShut
  {NoStop}%
\bibitem [{\citenamefont {Wel}\ \emph {et~al.}(2022)\citenamefont {Wel},
  \citenamefont {Allan}, \citenamefont {Keim},\ and\ \citenamefont
  {Caswell}}]{VanDerWel2022}%
  \BibitemOpen
  \bibfield  {author} {\bibinfo {author} {\bibnamefont {Wel}, \bibfnamefont
  {C.~V.~D.}}, \bibinfo {author} {\bibnamefont {Allan}, \bibfnamefont {D.}},
  \bibinfo {author} {\bibnamefont {Keim}, \bibfnamefont {N.}}, \ and\ \bibinfo
  {author} {\bibnamefont {Caswell}, \bibfnamefont {T.~A.}},\ }\href {\doibase
  10.5281/zenodo.4682814} {\enquote {\bibinfo {title} {Trackpy: Fast, flexible
  particle-tracking toolkit},}\ } (\bibinfo {year} {2022})\BibitemShut
  {NoStop}%
\bibitem [{\citenamefont {Wildenschild}\ and\ \citenamefont
  {Sheppard}(2013)}]{Wildenschild2013}%
  \BibitemOpen
  \bibfield  {author} {\bibinfo {author} {\bibnamefont {Wildenschild},
  \bibfnamefont {D.}}\ and\ \bibinfo {author} {\bibnamefont {Sheppard},
  \bibfnamefont {A.~P.}},\ }\bibfield  {title} {\enquote {\bibinfo {title}
  {X-ray imaging and analysis techniques for quantifying pore-scale structure
  and processes in subsurface porous medium systems},}\ }\href {\doibase
  10.1016/j.advwatres.2012.07.018} {\bibfield  {journal} {\bibinfo  {journal}
  {Advances in Water Resources}\ }\textbf {\bibinfo {volume} {51}},\ \bibinfo
  {pages} {217--246} (\bibinfo {year} {2013})},\ \bibinfo {note} {35th Year
  Anniversary Issue}\BibitemShut {NoStop}%
\bibitem [{\citenamefont {Ye}\ \emph {et~al.}(2019)\citenamefont {Ye},
  \citenamefont {Pan}, \citenamefont {Huang},\ and\ \citenamefont
  {Liu}}]{Ye2019}%
  \BibitemOpen
  \bibfield  {author} {\bibinfo {author} {\bibnamefont {Ye}, \bibfnamefont
  {T.}}, \bibinfo {author} {\bibnamefont {Pan}, \bibfnamefont {D.}}, \bibinfo
  {author} {\bibnamefont {Huang}, \bibfnamefont {C.}}, \ and\ \bibinfo {author}
  {\bibnamefont {Liu}, \bibfnamefont {M.}},\ }\bibfield  {title} {\enquote
  {\bibinfo {title} {Smoothed particle hydrodynamics (sph) for complex fluid
  flows: Recent developments in methodology and applications},}\ }\href
  {\doibase 10.1063/1.5068697} {\bibfield  {journal} {\bibinfo  {journal}
  {Physics of Fluids}\ }\textbf {\bibinfo {volume} {31}},\ \bibinfo {pages}
  {011301} (\bibinfo {year} {2019})}\BibitemShut {NoStop}%
\bibitem [{\citenamefont {Zarikos}\ \emph {et~al.}(2018)\citenamefont
  {Zarikos}, \citenamefont {Terzis}, \citenamefont {Hassanizadeh},\ and\
  \citenamefont {Weigand}}]{Zarikos2018a}%
  \BibitemOpen
  \bibfield  {author} {\bibinfo {author} {\bibnamefont {Zarikos}, \bibfnamefont
  {I.}}, \bibinfo {author} {\bibnamefont {Terzis}, \bibfnamefont {A.}},
  \bibinfo {author} {\bibnamefont {Hassanizadeh}, \bibfnamefont {S.~M.}}, \
  and\ \bibinfo {author} {\bibnamefont {Weigand}, \bibfnamefont {B.}},\
  }\bibfield  {title} {\enquote {\bibinfo {title} {Velocity distributions in
  trapped and mobilized non-wetting phase ganglia in porous media},}\ }\href
  {\doibase 10.1038/s41598-018-31639-4} {\bibfield  {journal} {\bibinfo
  {journal} {Scientific Reports}\ ,\ \bibinfo {pages} {1--11}} (\bibinfo {year}
  {2018})}\BibitemShut {NoStop}%
\bibitem [{\citenamefont {Zhang}\ \emph {et~al.}(2021)\citenamefont {Zhang},
  \citenamefont {Kaito}, \citenamefont {Hu}, \citenamefont {Patmonoaji},
  \citenamefont {Matsushita},\ and\ \citenamefont {Suekane}}]{Zhang2021}%
  \BibitemOpen
  \bibfield  {author} {\bibinfo {author} {\bibnamefont {Zhang}, \bibfnamefont
  {C.}}, \bibinfo {author} {\bibnamefont {Kaito}, \bibfnamefont {K.}}, \bibinfo
  {author} {\bibnamefont {Hu}, \bibfnamefont {Y.}}, \bibinfo {author}
  {\bibnamefont {Patmonoaji}, \bibfnamefont {A.}}, \bibinfo {author}
  {\bibnamefont {Matsushita}, \bibfnamefont {S.}}, \ and\ \bibinfo {author}
  {\bibnamefont {Suekane}, \bibfnamefont {T.}},\ }\bibfield  {title} {\enquote
  {\bibinfo {title} {Influence of stagnant zones on solute transport in
  heterogeneous porous media at the pore scale},}\ }\href {\doibase
  10.1063/5.0038133} {\bibfield  {journal} {\bibinfo  {journal} {Physics of
  Fluids}\ }\textbf {\bibinfo {volume} {33}},\ \bibinfo {pages} {036605}
  (\bibinfo {year} {2021})}\BibitemShut {NoStop}%
\bibitem [{\citenamefont {Zhao}\ \emph {et~al.}(2019)\citenamefont {Zhao},
  \citenamefont {MacMinn}, \citenamefont {Primkulov}, \citenamefont {Chen},
  \citenamefont {Valocchi}, \citenamefont {Zhao}, \citenamefont {Kang},
  \citenamefont {Bruning}, \citenamefont {McClure}, \citenamefont {Miller},
  \citenamefont {Fakhari}, \citenamefont {Bolster}, \citenamefont {Hiller},
  \citenamefont {Brinkmann}, \citenamefont {Cueto-Felgueroso}, \citenamefont
  {Cogswell}, \citenamefont {Verma}, \citenamefont {Prodanović}, \citenamefont
  {Maes}, \citenamefont {Geiger}, \citenamefont {Vassvik}, \citenamefont
  {Hansen}, \citenamefont {Segre}, \citenamefont {Holtzman}, \citenamefont
  {Yang}, \citenamefont {Yuan}, \citenamefont {Chareyre},\ and\ \citenamefont
  {Juanes}}]{Zhao2019}%
  \BibitemOpen
  \bibfield  {author} {\bibinfo {author} {\bibnamefont {Zhao}, \bibfnamefont
  {B.}}, \bibinfo {author} {\bibnamefont {MacMinn}, \bibfnamefont {C.~W.}},
  \bibinfo {author} {\bibnamefont {Primkulov}, \bibfnamefont {B.~K.}}, \bibinfo
  {author} {\bibnamefont {Chen}, \bibfnamefont {Y.}}, \bibinfo {author}
  {\bibnamefont {Valocchi}, \bibfnamefont {A.~J.}}, \bibinfo {author}
  {\bibnamefont {Zhao}, \bibfnamefont {J.}}, \bibinfo {author} {\bibnamefont
  {Kang}, \bibfnamefont {Q.}}, \bibinfo {author} {\bibnamefont {Bruning},
  \bibfnamefont {K.}}, \bibinfo {author} {\bibnamefont {McClure}, \bibfnamefont
  {J.~E.}}, \bibinfo {author} {\bibnamefont {Miller}, \bibfnamefont {C.~T.}},
  \bibinfo {author} {\bibnamefont {Fakhari}, \bibfnamefont {A.}}, \bibinfo
  {author} {\bibnamefont {Bolster}, \bibfnamefont {D.}}, \bibinfo {author}
  {\bibnamefont {Hiller}, \bibfnamefont {T.}}, \bibinfo {author} {\bibnamefont
  {Brinkmann}, \bibfnamefont {M.}}, \bibinfo {author} {\bibnamefont
  {Cueto-Felgueroso}, \bibfnamefont {L.}}, \bibinfo {author} {\bibnamefont
  {Cogswell}, \bibfnamefont {D.~A.}}, \bibinfo {author} {\bibnamefont {Verma},
  \bibfnamefont {R.}}, \bibinfo {author} {\bibnamefont {Prodanović},
  \bibfnamefont {M.}}, \bibinfo {author} {\bibnamefont {Maes}, \bibfnamefont
  {J.}}, \bibinfo {author} {\bibnamefont {Geiger}, \bibfnamefont {S.}},
  \bibinfo {author} {\bibnamefont {Vassvik}, \bibfnamefont {M.}}, \bibinfo
  {author} {\bibnamefont {Hansen}, \bibfnamefont {A.}}, \bibinfo {author}
  {\bibnamefont {Segre}, \bibfnamefont {E.}}, \bibinfo {author} {\bibnamefont
  {Holtzman}, \bibfnamefont {R.}}, \bibinfo {author} {\bibnamefont {Yang},
  \bibfnamefont {Z.}}, \bibinfo {author} {\bibnamefont {Yuan}, \bibfnamefont
  {C.}}, \bibinfo {author} {\bibnamefont {Chareyre}, \bibfnamefont {B.}}, \
  and\ \bibinfo {author} {\bibnamefont {Juanes}, \bibfnamefont {R.}},\
  }\bibfield  {title} {\enquote {\bibinfo {title} {Comprehensive comparison of
  pore-scale models for multiphase flow in porous media},}\ }\href {\doibase
  10.1073/pnas.1901619116} {\bibfield  {journal} {\bibinfo  {journal}
  {Proceedings of the National Academy of Sciences}\ }\textbf {\bibinfo
  {volume} {116}},\ \bibinfo {pages} {13799--13806} (\bibinfo {year}
  {2019})}\BibitemShut {NoStop}%
\end{thebibliography}%

\end{document}